\documentclass[sigconf]{acmart}

\usepackage{booktabs} 

\usepackage{balance}       
\usepackage{graphics}      
\usepackage{times}         
\usepackage{paralist}      
\usepackage{booktabs}      
\usepackage{subfigure}     

\usepackage{amsmath}
\usepackage{amssymb}
\usepackage{algorithm}
\usepackage{algorithmic}
\usepackage{xspace}

\newcommand{\spara}[1]{\smallskip\noindent{\bf #1}}

\newcommand{\News}{\textit{News}\xspace}
\newcommand{\OpEd}{\textit{Opinion}\xspace}

\newcommand{\topics}{\ensuremath{\mathcal{T}}\xspace}
\newcommand{\articles}{\ensuremath{\mathcal{D}}\xspace}

\newcommand{\nnd}[3]{\ensuremath{N_#1^{#2}(#3)}\xspace}

\newcommand{\relevance}[2]{\ensuremath{\operatorname{rel}(#1,#2)}\xspace}

\newcommand{\articlecumvisits}[2]{\ensuremath{V_{#1}(#2)}}
\newcommand{\topicvisits}[2]{\ensuremath{Y_{#1}(#2)}}

\newcommand{\articlecumvisitsest}[2]{\ensuremath{\widehat{V}_{#1}(#2)}}
\newcommand{\topicvisitsest}[2]{\ensuremath{\widehat{Y}_{#1}(#2)}}

\newcommand{\articlesim}[2]{\ensuremath{\operatorname{sim}_{\operatorname{cos}}(#1,#2)}}

\newcommand{\timelags}[1]{\ensuremath{\delta(#1)}}

\begin{document}
%

\title{To Post or Not to Post: Using Online Trends to Predict Popularity of Offline Content}
\subtitle{(Extended version of ACM HT'2018 paper)}
\author{Sofiane Abbar$^1$, Carlos Castillo$^2$, Antonio Sanfilippo$^3$}
\affiliation{%
  \institution{$^1$Qatar Computing Research Institute, HBKU. Doha, Qatar\\
  $^2$Universitat Pompeu Fabra. Barcelona, Catalunya, Spain\\
  $^3$Qatar Environment \& Energy Research Institute, HBKU. Doha, Qatar\\
  \{sabbar,asanfilippo\}@hbku.edu.qa, chato@acm.org
  }
}

\renewcommand{\shortauthors}{S. Abbar et al.}
\renewcommand{\shorttitle}{To Post or Not to Post: Popularity of Offline Content}
\begin{abstract}
Predicting the popularity of online content has attracted much attention in the past few years.
In news rooms, for instance, journalists and editors are keen to know, as soon as possible, the articles that will bring the most traffic into their website.
The relevant literature includes a number of approaches and algorithms to perform this forecasting. Most of the proposed methods require monitoring the popularity of content during some time after it is posted, before making any longer-term prediction.
In this paper, we propose a new approach for predicting the popularity of news articles before they go online. Our approach complements existing content-based methods, and is based on a number of observations regarding article similarity and topicality.
First, the popularity of a new article is correlated with the popularity of similar articles of recent publication. Second, the popularity of the new article is related to the recent historical popularity of its main topic.
Based on these observations, we use time series forecasting to predict the number of visits an article will receive. Our experiments, conducted on a real data collection of articles in an international news website, demonstrate the effectiveness and efficiency of the proposed method.

\end{abstract}

\maketitle

\section{Introduction}\label{sec:intro}
Monitoring the performance of news articles is a core task within any news media organization. The highly crowded news market, and the fast growth of online news platforms and applications in recent years, have pushed editors into a fierce competition for the attention of news readers.
Social media are changing the way people consume news~\cite{kwak:2010,pew2013trends}, but they still constitute a small portion of the overall online news traffic. For instance, Andrew Miller, Guardian News and Media CEO, said that social media all combined add up to around 10\% of their newspaper's traffic.\footnote{https://blog.twitter.com/2013/guardian-says-twitter-surpassing-other-social-media-for-breaking-news-traffic} 
Currently, editors focus on \textit{popularity} in terms of number of visits and visitors to news websites as the most important performance metric for news articles online.

Measuring popularity, however, is not sufficient. The ability to \textit{anticipate} online news popularity enables editorial teams to take tactical and strategic decisions to maximize the impact of their online content, such as promoting or demoting articles in their web pages, changing the wording of headers, allocating editorial resources to follow-up stories or features, designing promotional campaigns, etc. Given the high velocity of news, editors and journalists need to have popularity forecasts for news articles as early as possible after publishing the article---and ideally, even before that.

The research community has addressed the problem of predicting the popularity of news articles in several recent papers including \cite{Lerman:2010, Tsagkias:2010, Bandari:2012, Tatar:2012, Castillo:2014}. 
Most of the proposed techniques rely on early measurements of visits and visitors to news websites, and are based on the auto-correlation of the time series that describe ebbs and flows in news popularity.

For example, a common method introduced by~\citet{Szabo:2010} is based on the observation that in some websites, there is a strong linear relationship between log-transformed early popularity and log-transformed long-term popularity, with correlations as high as $r=0.9$. This result makes it possible to forecast the future popularity of an article based on its early observed popularity.
Generalizations of this method have emerged since, including~\cite{li2013popularity,rowe2011forecasting} and others.

Naturally, the quality of these forecasts is lower the earlier the predictions are made, both because there is less data available, and because the time span between prediction time and target time is longer.
Moreover, predictions made \textit{before} articles go online are desirable, as these predictions allow editorial teams to take news management decisions without having to wait for early popularity measurements. Approaches that can dispense with early popularity measurements have been explored through the development of predictive models that use features such as the words in the title of the article, e.g. \cite{yu_2011_predicting,lakkaraju_2013_reddit}. Our approach is complementary to such content-based methods, and provides a novel extension where topic popularity forecasts are used to improve news article popularity predictions.

\spara{Our contribution.}
We introduce a new method for early prediction of popularity of news articles that combines article topicality and article similarity.
We show that the \textit{popularity of a topic} (the total number of visits received by all articles on that topic) depends on the popularity of related topics, and describe how to use this dependency to improve topic popularity predictions.
Next, we show that the \textit{popularity of an article} depends on the popularity of recent articles similar to it, and on the popularity of its general primary topic, which we can predict with a high level of accuracy. Finally, we propose an extension of the emerging approach, where topic popularity forecasts are used to improve news article popularity predictions.
We explore two forecasting algorithms that exploit these observations, and test them on a large collection of news articles published by an international news organization over 18 months in 2013 and 2014. The ensuing results yield a mean average percentage error as low as 11\% demonstrating the efficacy of the approach in predicting news article popularity.  

The paper is organized as follows. First, we provide an overview of related work. Then, we provide a detailed description of the data used in our study and discuss some of their characteristics (Section~\ref{sec:data}). Next, we present two predictive models of topic popularity (Section~\ref{sec:topic-predictions}), and proceed with a discussion of article popularity prediction (Section~\ref{sec:article-predictions}). We conclude by summarizing the novelty an impact of this research and its future extensions.

\section{Related Work} \label{sec:related}

The increasing use of predictive models of online content popularity in the news industry has promoted the growth of the already significant interest in predictive models of online user behavior in the research community. For ease of exposition, we limit our review to research that is closely related to the study presented in this paper. 

{\bf Methods Based on Early Measurements.}
The success of the auto-correlation approach pioneered by \citet{Szabo:2010,szabo2012predicting} has encouraged many researchers to use early popularity measurements as predictors of future popularity. 
Predictive models of online popularity based on auto-correlation have been used by:
\citet{Jamali:2009} with reference to votes in Digg;
\citet{lee_2010_popularity} for comments to articles;
\citet{lerman_2010_news} for visits to articles;
\citet{kim_2011_temperature} for visits to blog posts;
\citet{tatar_2011_predicting} for comments on articles;
\citet{ruan_2012_prediction} for number of Twitter messages---``tweets'';
\citet{Pinto:2013} for views in YouTube, and
\citet{Ahmed:2013} for views in YouTube and Vimeo, and for votes in Digg.
Many of these works use content metadata, such as publication date, and in some cases information about the users who post this content (e.g.~\cite{Jamali:2009,ruan_2012_prediction}).
Closer to the topic of this paper, the number of postings received by an article in social media (e.g. Twitter or Facebook) has been shown to be useful to predict visits to the article~\cite{Castillo:2014,Hsieh:2013}.

Our approach differs from these auto-correlation approaches in two main regards. First, early popularity measurements are not needed to provide reliable popularity predictions, although they can be incorporated in the algorithm. Second, we introduce the use of cross-correlations among topics as an important factor to improve the accuracy of predictions for topics and articles popularity.

{\bf Topic-Based Methods.}
\citet{Bandari:2012} used information about the category of a news article (e.g. sports, politics, technology) together with information about the communication source, language subjectivity, and named entities present in the article to predict the popularity of news articles in social media, prior to their publication. Scores for the communication source and category were computed as the average number of tweets per article for each news source and each category. The named entity score was computed in the same way, except that only the highest scoring named entity is selected among those appearing in each article (other variations were also tested). The prediction was done using linear regression resulting in $r^2=0.34$.
\citet{tatar_2011_predicting,Tatar:2012} predicted the number of comments to articles on a large news website. The prediction was based on linear regression using early data measurements. Articles in this website are separated into categories (world, sports, economy, etc.) Interestingly, a per-category model showed no improvements over a generic model that was oblivious to the category of an article.

We exploit the insight emerging from these methods that using the popularity of a topic in the distant past may not be the best predictor of future success, and provide a methodology for establishing the ideal time window.

{\bf Methods Based on Keywords}
Some predictive methods utilized a selection of keywords present in an article or headline as features for the popularity prediction model \cite{tsagkias2009predicting,lakkaraju_2011_attention,Berger:2012,lakkaraju_2013_reddit}. The intuition of this approach is that some of these keywords may be important for stylistic reasons (e.g. words such as as ``shocking'' or ``dramatic'' may attract more clicks), or because they refer to prominent people or powerful countries, which are important news values~\cite{galtung_1965_foreign}. 
For instance, authors of \cite{tsagkias2009predicting} have studied the prediction of comments on news articles, using metadata about the articles (e.g. publication date), the number of articles posted at the same time, the number of similar articles posted at the same time in other sources, and named entities mentioned in the article. 
Others \cite{Berger:2012} looked at articles that make it into the ``most emailed'' list of a large online newspaper, \textit{The New York Times}. Their focus was on two aspects of the articles' sentiment: polarity (``valence'') and emotionality (``arousal''), obtained through automated sentiment analysis. 
In \cite{lakkaraju_2013_reddit}, authors have also measured the popularity (positive minus negative votes) of an image \textit{re-posts} for different communities in a popular content sharing site, \textit{Reddit}. 
Results ranged from $r^2=0.36$ to $r^2=0.49$.
Finally, authors of \cite{lakkaraju_2011_attention} has focused on Facebook data to predict the number of comments a post will get. 
Support vector regression (SVR) was used to create predictive models achieving a correlation of $r^2=0.54$ with observed values.
Our approach differs from and is complementary to the approaches reviewed in this section, in that our approach relies on articles' content, topics, and ads.

{\bf Social Cascade Predictions.}
The prediction of information cascades in social networks has been an extremely active topic in recent years, particularly at the macroscopic level (i.e. how many nodes will be activated by a cascade), e.g.~\cite{Cheng:2014, li2013popularity, myers_2012_external, huang_2012_predicting} and many others. 
However, a setting in which social influence occurs may bring a high degree of unpredictability. \citet{Salganik:2006} claim that with social the popularity of an item is not an aggregate of individual preferences and therefore cannot be predicted even with perfect information: ``there are inherent limits on the predictability of outcomes, irrespective of how much skill or information one has''~\cite{Salganik:2006}.
These methods offer useful insights, but are not directly relevant to the problem focus of this study.


\section{Dataset}\label{sec:data}

In this section, we discuss the data used in our study. We describe how we generated the dataset from the source data (Section~\ref{subsec:data-preproc}), provide some insights on the intrinsic features that characterize the popularity of articles within our collection (Section~\ref{subsec:data-pop-distr}) and measure their effective life-span (Section~\ref{subsec:data-lifespan}). 

\subsection{Dataset Generation}\label{subsec:data-preproc}
We use data provided by \textit{(omitted for double-blind review)}, a large international news network operating multiple television channels and websites. We harvested articles from the English version of this website, which has millions of visits per month. The data covers a time span from September 2012 through April 2014. 
Our collection comprises two types of articles: \News and \OpEd. The first category refers to breaking news, reporting events and issues happening in different locations around the world. The second category refers to opinions and features contributed by named writers to present their opinion or analysis of a topic of public interest.
The collection consists of a sample of 8,065 \News articles and 4,357 \OpEd articles. Each article includes: title, content, and publication date.

For each article, we also retrieved a time series of the number of visits the article gets after its publication.
These time series are captured thanks to a large scale real-time process that records activities by single users per session on a minute by minute basis.

\subsection{Distribution of Visits}\label{subsec:data-pop-distr}

\begin{figure}[h!]
\centering\includegraphics[width=1.0\columnwidth]{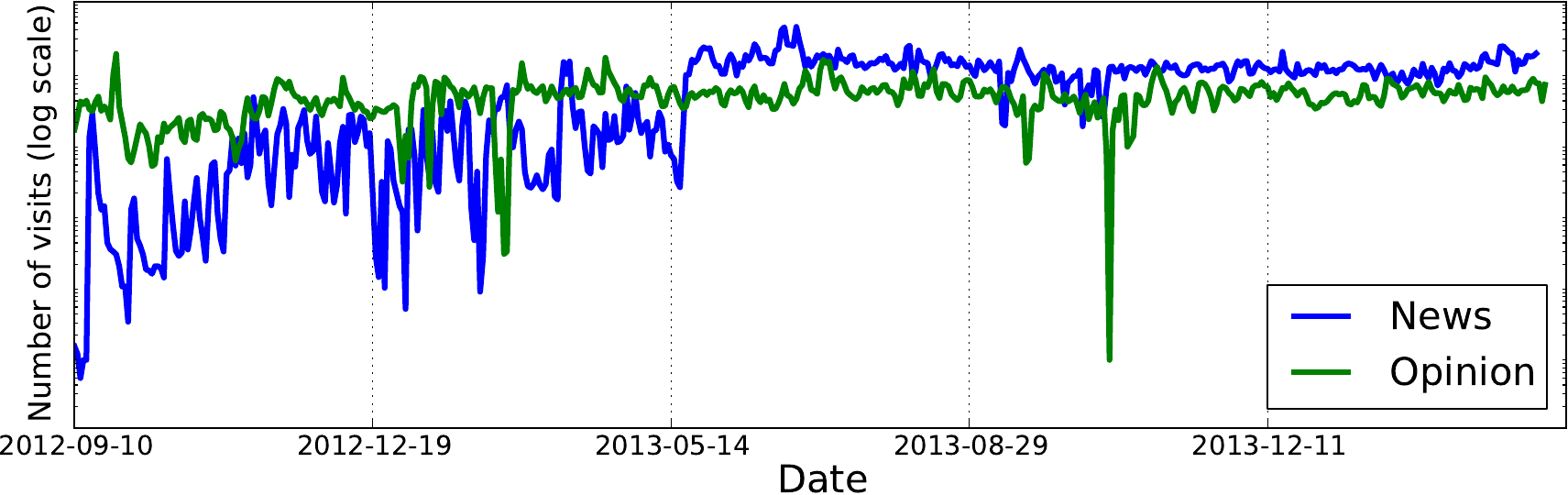}
\caption{Time series of total volume of visits to \News and \OpEd articles, during a span of a year and half. Note the Y-axis is in log scale.}
\label{fig:news-oped-series-6m}
\end{figure}

The overall time series of visits for the two sets of articles is shown in Figure~\ref{fig:news-oped-series-6m}. 
%
As shown in Figure ~\ref{fig:news-oped-series-6m}, the time series for \News is more variable than that for \OpEd articles. This difference reflects the more ephemeral nature of breaking news as compared to \OpEd articles, and is corroborated by a shorter shelf-life for breaking news, as shown in Figure~\ref{fig:life-age-cutoff}.

\begin{figure}[h!]
\centering\includegraphics[width=.7\columnwidth]{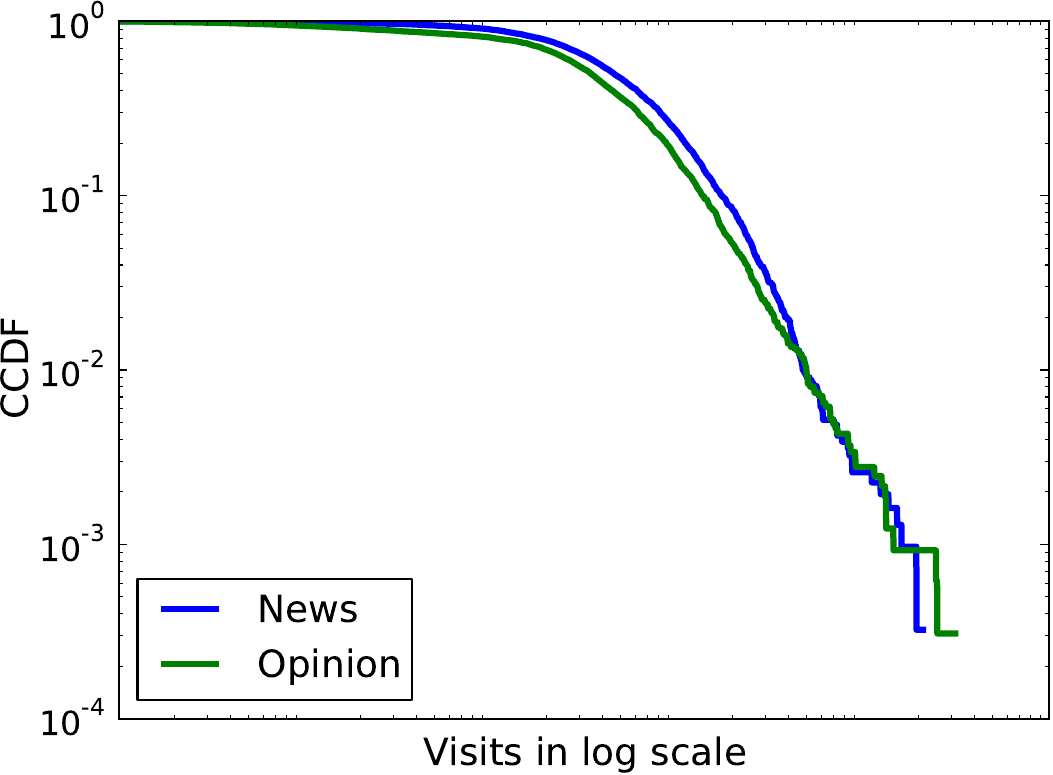}
\caption{Distribution of number of visits to \News and \OpEd articles in the first 30 days after their publication.}
\label{fig:dist-visits}
\end{figure}

The average number of visits for each article is in the order of a few thousands, but there are some articles that have hundreds of thousands of visits, and others that have only a few hundred.\footnote{Due to our legal agreement with the data provider, including the business-competitive nature of this data, we are not allowed to provide exact figures that can be used to estimate the total traffic to the website.} 
Figure~\ref{fig:dist-visits} shows the complementary cumulative distribution function (CCDF) of the number of visits articles receive in the first 30 days after publication. The popularity distribution is heavy tailed, which is in agreement with observations in e.g.~\cite{Tsagkias:2010,Lerman:2010,Castillo:2014}. 

\subsection{Shelf-Life of Articles: an Elusive Concept}\label{subsec:data-lifespan}

\begin{figure}[h]
\centering\includegraphics[width=.7\columnwidth]{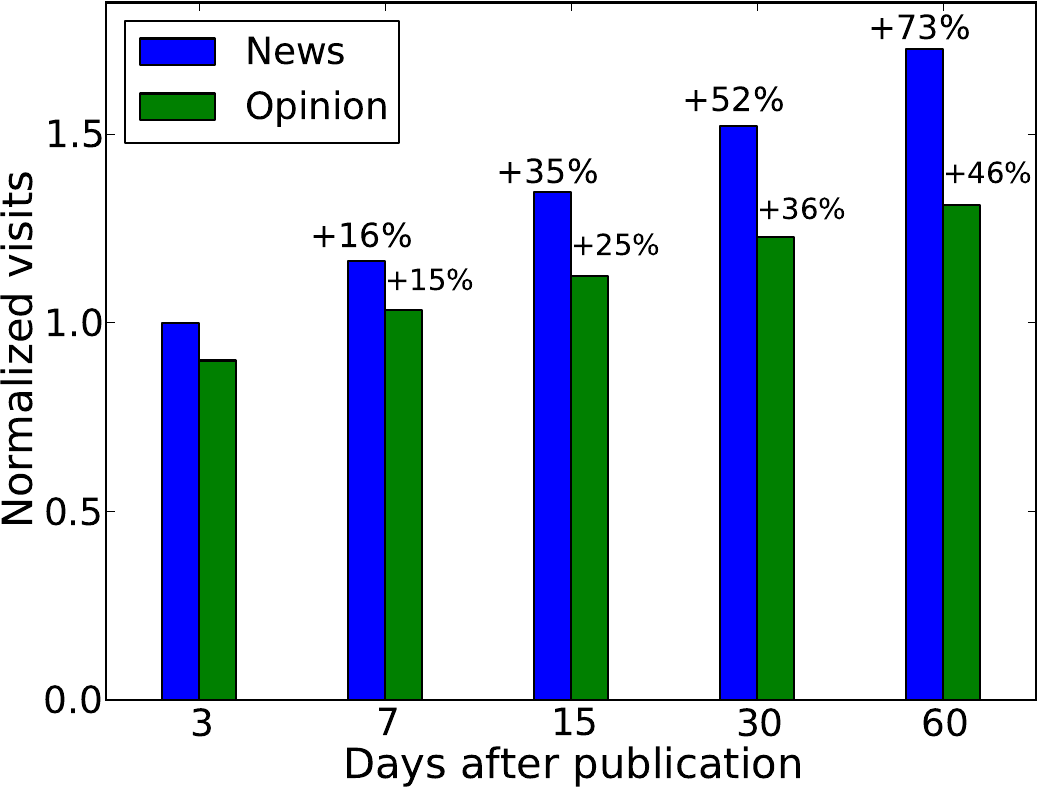}
\caption{Number of visits to average \News and \OpEd articles, expressed in terms of the number of visits to a \News article after 3 days, which is defined as 1.0 for normalization purposes.}
\label{fig:visits-growth}
\end{figure}

Readers' interest in news articles decreases sharply as time passes (as observed e.g. in~\cite{Tsagkias:2010,tatar_2011_predicting, Castillo:2014,Tatar:2012}). For example, 48\% of the visits for an average \News article in our dataset, over a 30-day period, occurs within the first three days, as shown in Figure~\ref{fig:visits-growth}.

\begin{figure}[h]
\centering
\subfigure[Percentage of visit for \News and \OpEd articles, within a 30 day time horizon.\label{fig:life-30}]{\includegraphics[width=.7\columnwidth]{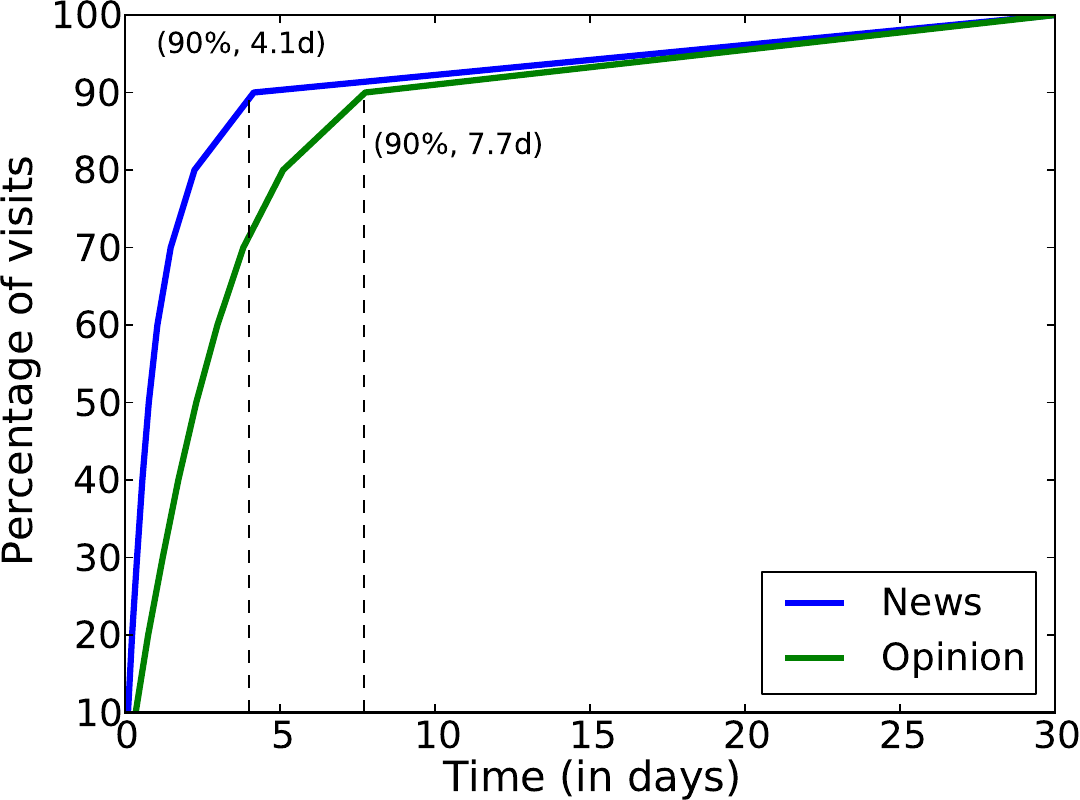}}
\subfigure[Shelf-life at 90\% for \News and \OpEd articles with progressively longer time spans.\label{fig:life-age-cutoff}]{\includegraphics[width=.7\columnwidth]{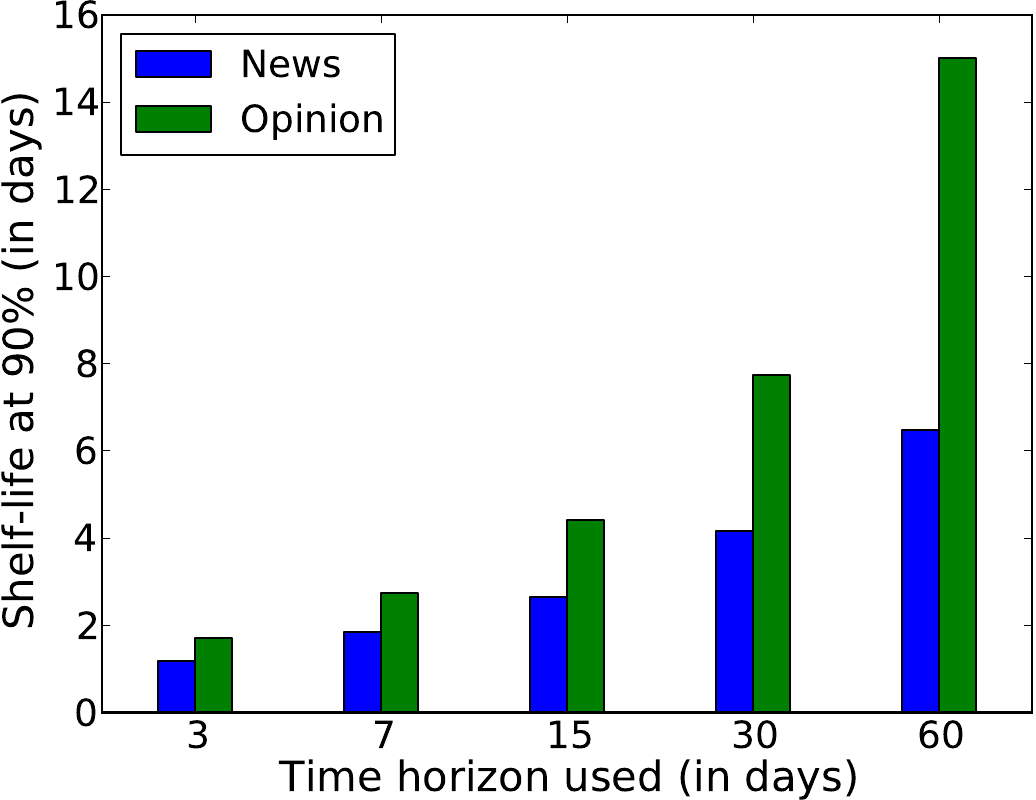}}
\caption{Computation of shelf-life at 90\% for \News and \OpEd articles. Top: within a 30 day time horizon. Bottom: using a varying time horizon.}
\label{fig:shelf-life}
\end{figure}

To measure the shelf-life of articles, we follow~\cite{dezso_2006_dynamics, Castillo:2014, bitly_2011_halflife, bitly_2012_halflife} and compute the time required for an article to reach a certain percentage of its visits. Specifically, we use the notion of \textit{shelf-life at 90\%}~\cite{Castillo:2014}, which is the time an article requires to accumulate 90\% of the visits it will receive in its lifetime.
Figure~\ref{fig:life-30} depicts the shelf-life at 90\% for \News (4.1 days on average) and \OpEd (7.7 days on average).
We observe that visits are more concentrated around the publication date in \News articles as compared to \OpEd articles, where visits are more spread-out in time. This is probably due to the fact that \OpEd articles are usually discussed longer and are not posted in reaction to immediate events as \News articles tend to be. 

In our 18-months dataset, articles posted online continued receiving visits long after their date of publication. New visitors may be directed to the article page as the result of a search engine query, through hyperlinks in more recent articles, or by consulting one of several thematic indexes on news websites.\footnote{This is in contrast with other measurements such as those for Twitter postings. People rarely tweet ``old'' articles on Twitter, so one can define the ``longevity'' of a news item as simply the time between the first and last tweet referring to the article~\cite{Hsieh:2013}.}
This makes defining an absolute \textit{shelf-life} difficult: it depends on the time horizon used to compute it, as Figure~\ref{fig:life-age-cutoff} shows.
In general, we observe a monotonic decreasing trend of the proportion between the shelf-life at 90\% and the time horizon used to compute it. While the shelf-life at 90\% accounts for no less than 39\% of the time for \News articles and 57\% for \OpEd articles when the time horizon equals 7 days, these proportions decrease to 10\% and 25\%  respectively when the time horizon is extended to 60 days. 

There are interesting differences between \News and \OpEd articles, such as the longer shelf-life for \OpEd articles, that may have an impact on the prediction of article popularity. In the remainder of the paper, we threat the two kind of articles as a single class of content type. We can easily obtain separate results for the two types of articles since the prediction method is the same, and we plan to do so in an extended version of this paper. 

\newcommand{\tfidf}{\ensuremath{\mathrm{tf} \cdot \mathrm{idf}}\xspace}

\begin{table}[t]
\caption{Summary of Notation}
\label{tbl:notation}
\centering\begin{tabular}{cp{2.65in}}\toprule
$\topics$       & Set of topics\\
$u \in \topics$ & A topic\\
$k = |\topics|$ & Number of topics\\
\midrule
$\articles$ & Set of articles \\
$a \in \articles$ & An article\\
$n = |\articles|$ & Number of articles \\
$\articlesim{a}{b}$       & Similarity of articles $a$ and $b$ \\
$\relevance{a}{u}$ & Relevance score of article $a$ to topic $u$ \\
$u_a$ & The most relevant topic for $a$: $\operatorname{argmax}_{u \in \topics} \relevance{a}{u}$ \\
\midrule
$t_a$    & Publication date of article $a$ \\
$\delta$ & Time lag expressed in days \\
$\timelags{t}$ & Set of time lags: $\{ t - \delta, t - \delta + 1, \dots, t - 1 \}$ \\
$\nnd{\theta}{\delta}{a}$ & Set of articles with $\articlesim{a}{\cdot} \ge \theta$, and published on date $t_a - \delta$\\
\midrule
$\articlecumvisits{a}{t}$ & For $t \ge t_a$, cumulative number of visits received by article $a$ on days $t_a, t_a + 1, \dots, t$ \\
$\topicvisits{u}{t}$    & Total number of visits to topic $u$ received on day $t$\\
\bottomrule
\end{tabular}
\end{table}

\section{Predicting topic volume}\label{sec:topic-predictions}

The first task we describe is the prediction of the total volume of visits to a topic $u$, i.e. the sum of the visits of all articles that have the topic $u$ as the main topic.
We apply Latent Dirichlet Allocation (LDA) as topic modeling method (Section~\ref{subsec:topic-lda}), determine the optimal number of topics using supervised classification (Section~\ref{subsec:topic-number}), describe the forecasting methods we use for topic volume prediction (Section~\ref{subsec:topic-predictions}), and discuss their application to our dataset and the ensuing results (Section~\ref{subsec:topic-results}).

\subsection{Modeling Method for Topics: LDA}\label{subsec:topic-lda}

We use Latent Dirichlet allocation algorithm (LDA) to uncover the topics in our collection of articles~\cite{Blei:2003}. LDA is a probabilistic generative method that uses a Bayesian network to discover a set of latent topics \topics from a set of documents \articles. 
To prepare our articles for LDA, we first concatenate the title of the article with its body, then remove stop words, and stem the remaining words using the stemmer implementation by Paice and Husk, also known as the Lancaster stemming algorithm~\cite{Paice:1990}. 

LDA outputs the probability that an article $a \in \articles$ is about a topic $u \in \topics$, which we denote as $\relevance{a}{u}$. This and other notation used throughout this paper are summarized on Table~\ref{tbl:notation}.

\subsection{Determining the Number of Topics}\label{subsec:topic-number}

Like many methods used for identifying latent topics in documents, including non-negative matrix factorization (NMF)~\cite{Lee:2000} and Probabilistic Latent Semantic Analysis (pLSA)~\cite{Hofmann:1999}, LDA assumes the number of topics $k$ is known in advance. However, determining the optimal number of topics remains an open research question~\cite{Arun:2010}.
This choice is critical for our application, because topic volume prediction is sensitive to the number of topics $k$ selected (see below).
Empirically, if we use a small number of topics, LDA returns broad topics such as politics, sports and armed conflicts. But if we request a large number of topics, LDA creates specialized topics around specific stories or journalistic \textit{beats}, such as the US elections, the Egyptian elections, the Syrian conflict, and politics in Latin America.

We use supervised classification to find the ``appropriate'' number of topics $k^*$. The intuition is that $k^*$ topics should yield a partition of the documents in the dataset that can be accurately recognized by a classifier trained on $k^*$ classes of documents, each class corresponding to one of the selected $k^*$ topics. 
First, we run LDA with different number of topics $(k \in \{10, 20 , \dots, 100 \})$. Let $\topics^{(k)}$ be the topic set produced by LDA for each value of $k$.
For each set of topics, we label every article $a \in \articles$ with its primary topic $u_a^{(k)}$ such that $u_a^{(k)} = \operatorname{argmax}_{u \in \topics^{(k)}} \relevance{a}{u}$.

Then, we select 80\% of the entire collection of labeled articles for training and the remaining 20\% for testing. We train a Multinomial Naive Bayes classifier (MNB)~\cite{McCallum:1998} using the training data, and evaluate the classifier on the test data. The selected feature space is defined as the \tfidf scores of stems within each article. The classification quality achieved by the MNB is measured in terms of precision, recall, and $F_1$ score (the harmonic mean of precision and recall).

\begin{figure}[h]
\centering \includegraphics[width=.8\columnwidth]{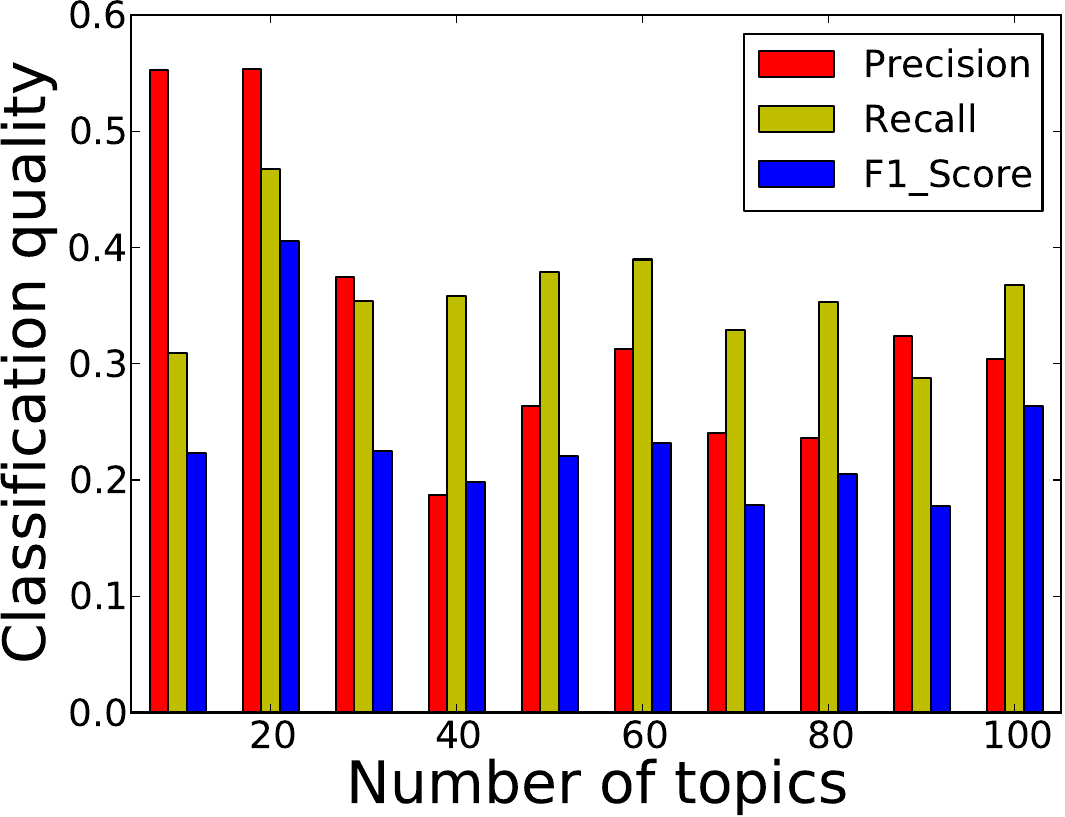}
\caption{Supervised classification quality for different numbers of topics obtained using LDA. We observe the maximum $F_1$ score at about 20 topics.}
\label{fig:lda-num-topics}
\end{figure}

Results for varying number of topics $k$ are reported in Figure~\ref{fig:lda-num-topics}. While the precision of the classifier is almost the same at $k=10$ and $k=20$, the recall and $F_1$ scores are maximized for $k=20$.
Based on these experiments, we select $k^*=20$ as the ideal number of topics to forecast topic volume for our dataset.

We note that the number of topics that yields the best classification model for a dataset is sensitive to the number and timespan of the articles in the dataset. In general, we have observed that as datasets get smaller so does the number of topics needed to yield the best possible classification model for the dataset. For example, 10 topics yield a better classification model than 20 topics do for a collection of about 1/3 of the articles contained in our dataset.

\subsection{Topic Popularity Prediction Methods}\label{subsec:topic-predictions}

We use a machine learning approach to forecasting, where each training sample is a pair $\langle \vec{x},y \rangle$, where $\vec{x} \in \mathbb{R}^n$ is an input vector of features for the time-series class to be learned, and $y \in \mathbb{R}$ its associated value. The aim of the machine learning algorithm is to find a function that for each $\vec{x}_i$ in the training dataset approximates its value $y_i$ as close as possible. The resulting function is then used to predict values n-steps ahead of the time series data used for training. We compare results from two algorithms, one based on linear regression (LR) and the other on support vector regression (SVR).

\spara{Linear Regression (LR).}
Within a linear regression approach for forecasting method~\cite{rowe2011forecasting, li2013popularity}, $\topicvisits{u}{t}$, the total number of visits to articles in topic $u$ at time $t$, is given by

\begin{equation}
\topicvisits{u}{t} = \alpha + \sum_{i \in \timelags{t}} \beta_i \topicvisits{u}{i} + \varepsilon
\end{equation}

Where $\alpha$ and $\beta_i$ are coefficients of the linear regression, $\varepsilon$ is a residual term, and $\timelags{t}$ is the set of time lags $\{ t - \delta, t - \delta + 1, \dots, t - 1 \}$.
In a more general version, we assume that the volume of visits of a topic depends not only on that topic (due to auto-correlation) but on all the other topics (due to cross-correlations):

\begin{equation}
\topicvisits{u}{t} = \alpha + \sum_{i \in \timelags{t}} \sum_{v \in \topics} \beta_{i,j} \topicvisits{v}{i} + \varepsilon
\end{equation}

\spara{Support Vector Regression (SVR).}
Within a Support Vector Regression (SVR) approach to time series forecasting \cite{muller1997predicting}, the prediction function is given by the formula:
\begin{equation}
\topicvisits{u}{t} = \vec{w} \cdot \vec{x}_u(t) + b
\end{equation}
where $\vec{w}$ is the weight vector, i.e. a linear combination of training patterns that supports the regression function,
$\vec{x}_u(t)$ is the vector containing the input features available at time $t$ (this is a vector containing all \topicvisits{v}{i} for $v \in \topics$ and $i \in \timelags{t}$),
and $b$ is the bias, i.e. an average over \textit{marginal vectors}, which are weight vectors that lie within the margins set by the loss function (see below).

The objective of SVR regression is to learn the weight vector $\vec{w}$ that has the smallest possible length so as to avoid over-fitting. To ease the regression task, a given margin of deviation $\varepsilon$ is allowed with no penalty, and a given margin $\xi$ is specified where deviation is allowed with increasing penalty. The length of the weight vector $\vec{w}$ is obtained by minimizing the loss function
\begin{equation*}
\frac{1}{2} ||\vec{w}||^2 + C \sum_{i=1}^n ( \xi_i + \xi_i^*)
\end{equation*}
subject to the constraints:
\begin{eqnarray*}
\topicvisits{u}{t} - ( \vec{w} \cdot \vec{x}_u(t) + b ) & \le & \varepsilon + \xi_i \mathrm{~~~~~or}\\
\topicvisits{u}{t} - ( \vec{w} \cdot \vec{x}_u(t) + b ) & \ge & -\varepsilon - \xi_i^*~,\\
\mathrm{with}~~\xi_i, \xi_i^* & \ge & 0~.
\end{eqnarray*}

\noindent The solution is given by the equation
\begin{equation*}
\topicvisits{u}{t} = \sum_{i=1}^n (\alpha_i - \alpha_i^*) (\vec{w} \cdot \vec{x}_u(t)) + b
\end{equation*}
where $\alpha_i$ and $\alpha_i^*$ are Lagrange multipliers---see \cite{smola2004tutorial} for details. 

\spara{Feature selection.}
There are numerous input variables $\topicvisits{v}{i}, i \in \timelags{t}, v \in \topics$, a total of $\delta |\topics|$, which can be relatively large compared to the number of observations. This may lead to over-fitting, so a topic selection method could in principle lead to better results---indeed we show in the next section that it is the case.
We apply a feature selection in which we select for each topic $u$ the topics that are most correlated with $u$ among the set of topics. Concretely, instead of using as input variables all $\topicvisits{v}{i}$ with $v \in \topics$, we select only the $s$ topics that have the largest cross-correlation with topic $u$ (in practice this includes the topic $u$ itself).

\subsection{Topic Volume Prediction Results}\label{subsec:topic-results}

We use Pearson's correlation ($r^2$) to measure auto-correlations and cross-correlations between topics, and Mean Absolute Percentage Error (MAPE) to evaluate forecasting results. MAPE is one of the most common measures of forecast error~\cite{armstrong1992error}. It expresses the error of the forecasted time series as a percentage:
\begin{equation}
\frac{1}{n}\sum_{i=1}^n \frac{|\topicvisits{u}{t} - \topicvisitsest{u}{t}|}{\topicvisits{u}{t}} \times 100
\end{equation}
where $\topicvisits{u}{t}$ and $\topicvisitsest{u}{t}$ are respectively the observed and forecasted values for topic $u$ at time $t$. When there is a perfect fit, MAPE is 0\%. There is no upper bound on the lack of fit.

\spara{Topic Volume Auto-Correlation.}
We first verify that the topics we determine are not only coherent in terms of content (as shown in the previous section), but also uncover auto-correlations in the time series of topic volume. This auto-correlation means that, for instance, a topic that was popular yesterday (or $\delta$ days ago) is likely to be popular today.
Specifically, we compute the correlation of each time series of total topic volume $\topicvisits{u}{t}$ with a $\delta$-shifted version of it $\topicvisits{u}{t-\delta}$. We varied $\delta$ from 1 day to 7 days. The average auto-correlation across topics in $\topics$ is shown in Table~\ref{tab:topic-autocorr}.

\begin{table}[h!]
\caption{Auto-correlation with lag $\delta$: correlation of the total volume of visits to a topic on a given day, with its total volume $\delta$ days before.}
\begin{center}
{\small
\begin{tabular}{cccccccc}\hline
     & $\delta=1$ & $\delta=2$ & $\delta=3$ &$\delta=4$ & $\delta=5$ & $\delta=6$ & $\delta=7$ \\\hline
$r^2$  & 0.70 & 0.52 & 0.43 & 0.36 & 0.32 & 0.30 & 0.27 \\\hline
\end{tabular}}
\end{center}
\label{tab:topic-autocorr}
\end{table}

Unsurprisingly, Table~\ref{tab:topic-autocorr} shows that topics are strongly auto-correl\-ated at small time lags. For instance, a correlation of 0.7 is observed between popularity scores picked within one day interval ($\delta=1$). This means that if a topic was highly popular yesterday, then it is highly likely that it will be popular today. The auto-correlation decreases as the time lag increases.

\spara{Impact of Feature Selection and LR vs SVR.}
We next run an experiment to test the feature selection method and to compare LR and SVR.
We present the results using features up to a time lag of $\delta=3$ days (results with time lags of 2, 4, and 5 days are basically equivalent). Given that we have 20 topics, this yields a total of $20\delta = 60$ variables. When applying feature selection, we select for each topic $u$ the top $s=4$ topics whose volumes are most correlated to $u$ (in terms of $r^2$), yielding a total of 12 variables.

\begin{figure}[h]
\centering
\subfigure[All features\label{fig:mlv-svr-weka-60}]{\includegraphics[width=.45\columnwidth]{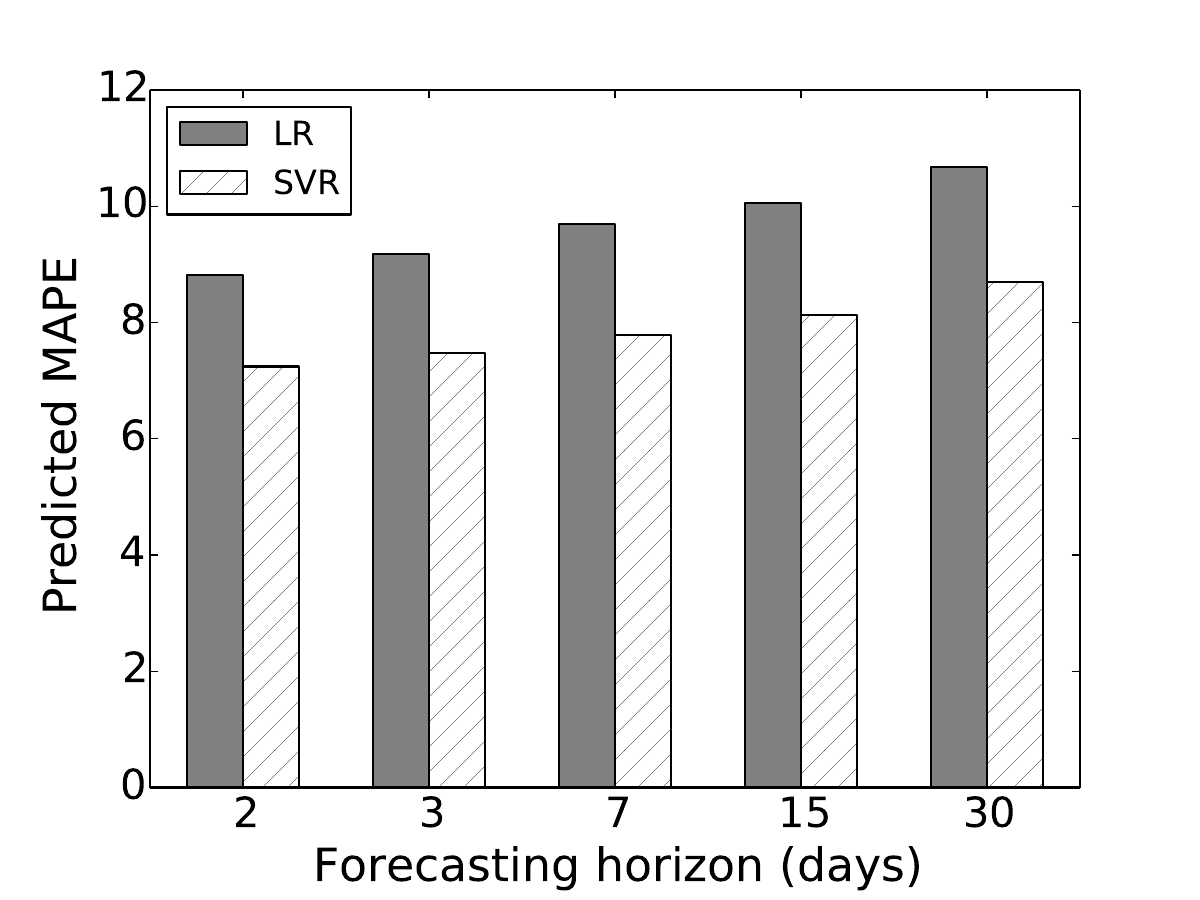}}
\subfigure[Top 12 features (4 topics)\label{fig:mlv-svr-weka-10}]{\includegraphics[width=.45\columnwidth]{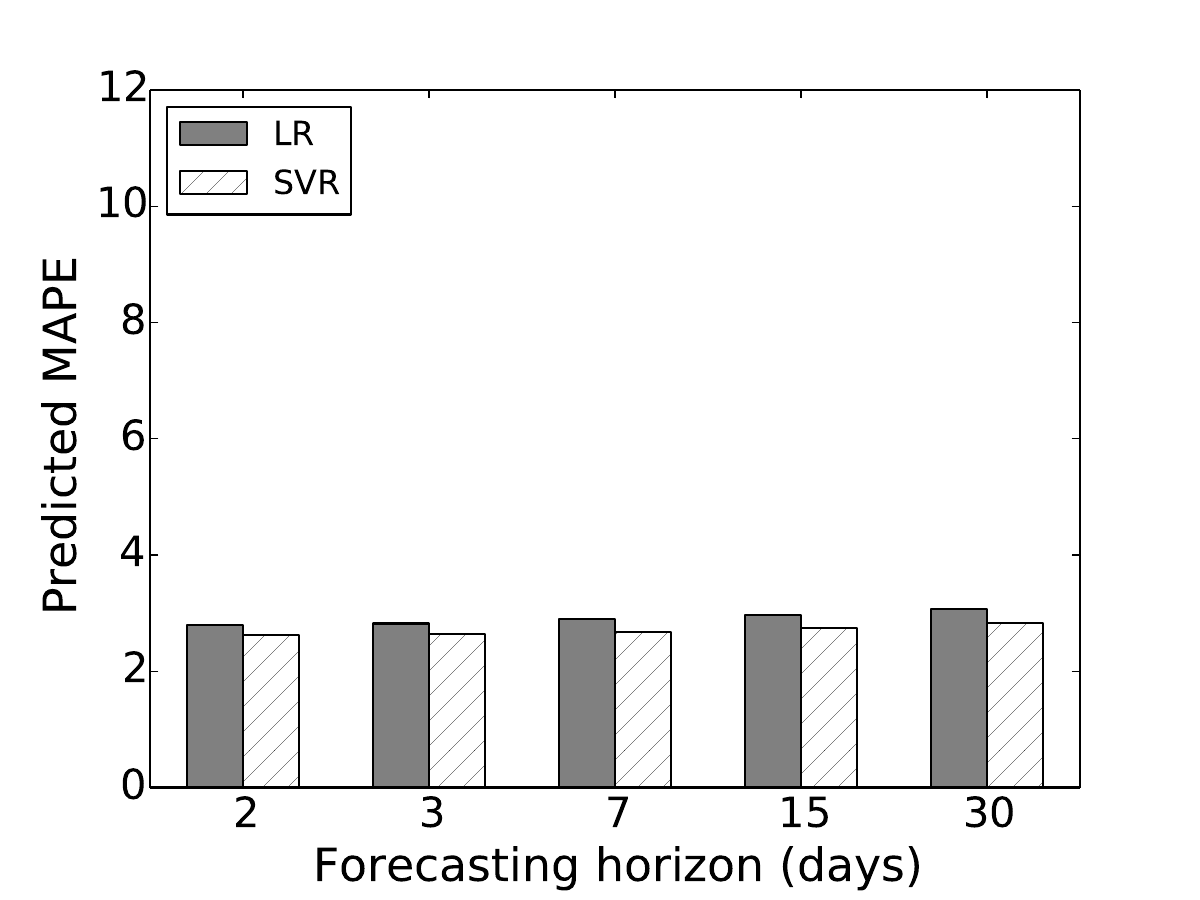}}
\caption{Comparison of prediction accuracy in terms of MAPE without feature selection (left) and with feature selection (right), using LR and SVR.}
\label{fig:mlv-svr}
\end{figure}

We train on a sliding window of 50 days (we show the impact of the time window size next), meaning that predictions for articles posted on day $t$, are done with a model trained on data from the days between $t-50$ and $t-1$.
To evaluate each method, we predict the topic volume for every topic at 2, 3, 7, 15, and 30 steps (days) ahead.
We report the achieved MAPE scores averaged across topics, comparing the prediction error obtained using all 60 features, shown in Figure~\ref{fig:mlv-svr-weka-60}, with the prediction error using the subset of 12 features, shown in Figure~\ref{fig:mlv-svr-weka-10}.

We make the following observations from Figure~\ref{fig:mlv-svr}. First, as expected the more steps ahead we try to forecast, the more errors we make. Second, the SVR method yields better MAPE results, particularly when no feature selection is applied. Third, and more importantly, feature selection dramatically increases the accuracy of this method, reducing MAPE significantly.

\spara{Determining the size of the training window.}
We now address the selection of the appropriate size of the time-window for training. In general, the size of the training set impacts the results of any machine learning algorithm. This is particularly true in the case of time series forecasting. A larger training window means more data is used for training, but \textit{if the underlying model changes over time}, then incorporating training data that is too old may actually be counterproductive.
The number of time lags $\delta$ to use is another important parameter. A larger $\delta$ means more variables are used for the prediction, which may lead to over-fitting.

We train our prediction models with training windows of different sizes and different time lag values. We vary the sliding training window size to take values in $\{3, 5, 7, 10, 20, 30, 40, 50, 60\}$, and the lag $\delta \in \{2, 3, 4, 5\}$, both values expressed in days. As before, we apply feature selection keeping the 4 topics most correlated to each topic.

\begin{figure}[h]
\subfigure[LR \label{fig:train-window-lr}]{\includegraphics[width=.48\columnwidth]{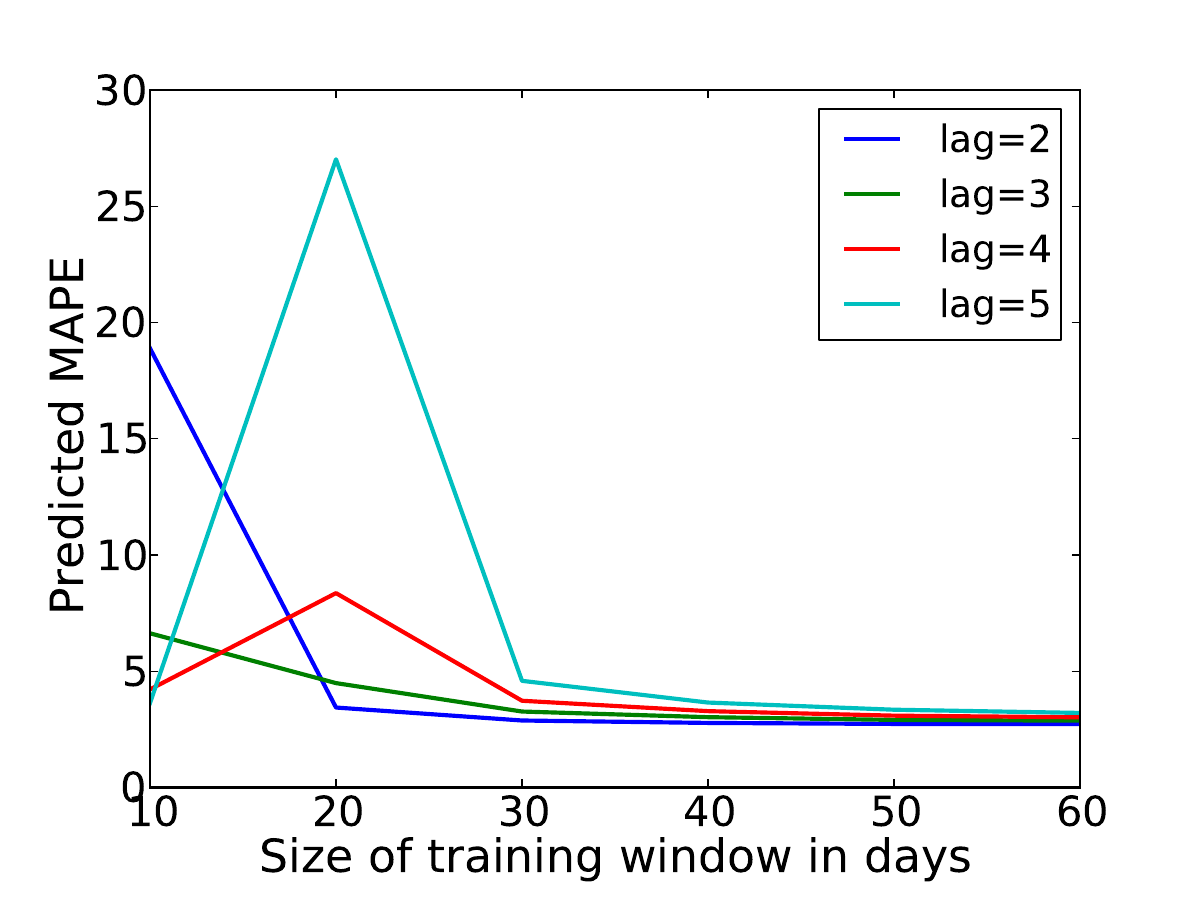}}
\subfigure[SVR \label{fig:train-window-svr}]{\includegraphics[width=.48\columnwidth]{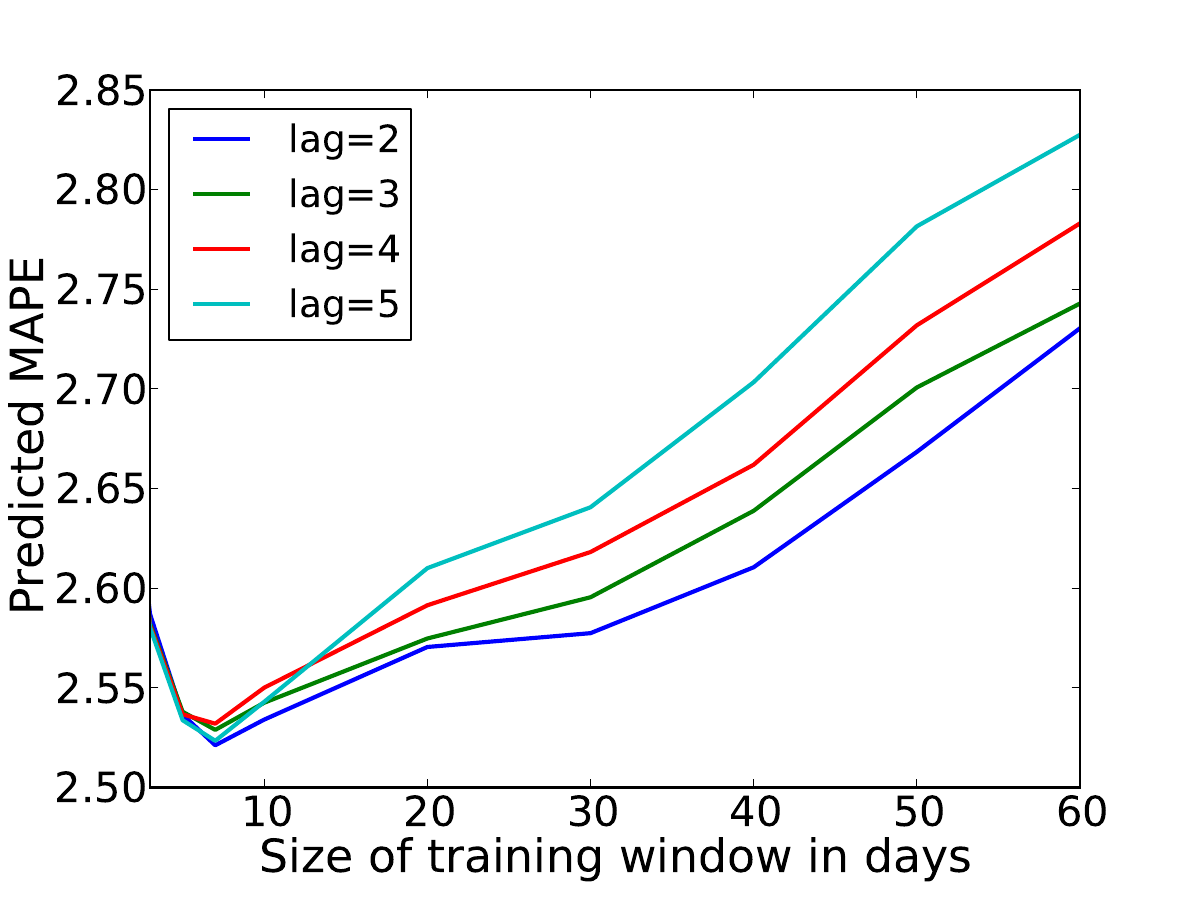}}
\caption{Comparison of LR and SVR in terms of MAPE, for different sizes of training window and lags. Each point is an average of the scores obtained across topics and steps-ahead.}
\label{fig:train-window-size}
\end{figure}

Figure~\ref{fig:train-window-size} reports the average MAPE scores computed for different values of time lags and sizes of the training set. Each reported MAPE value is the average of scores achieved at predicting different steps-ahead (2, 3, 7, 15, and 30).
Linear regression (LR) results are shown in Figure~\ref{fig:train-window-lr}. A high variation of MAPE scores is observed for small sizes of the training set ($\leq 30$) before the scores stabilizes starting from training sets of size 50.
Support vector regression (SVR) is shown in Figure~\ref{fig:train-window-svr} and it shows a different behavior.
First, it achieves much lower MAPE scores compared to those of LR, for all the values of the training set size we consider.
Second, with SVR the ideal size of the training window is achieved at 7 days, and thereafter, adding more observations increases the error rate.
Finally, adding more lags (larger $\delta$) also increases the error rate.

To summarize, the best topic prediction model we find is SVR with feature selection, a training window size of 7 days, and $\delta=2$ or $\delta=3$ as time lags.

\begin{figure*}
\centering
\subfigure[Performance of nearest-neighbors method NN (continuous line) vs. topic-based method T (dashed line)\label{fig:article-pred-NN}]{\includegraphics[width=.31\textwidth]{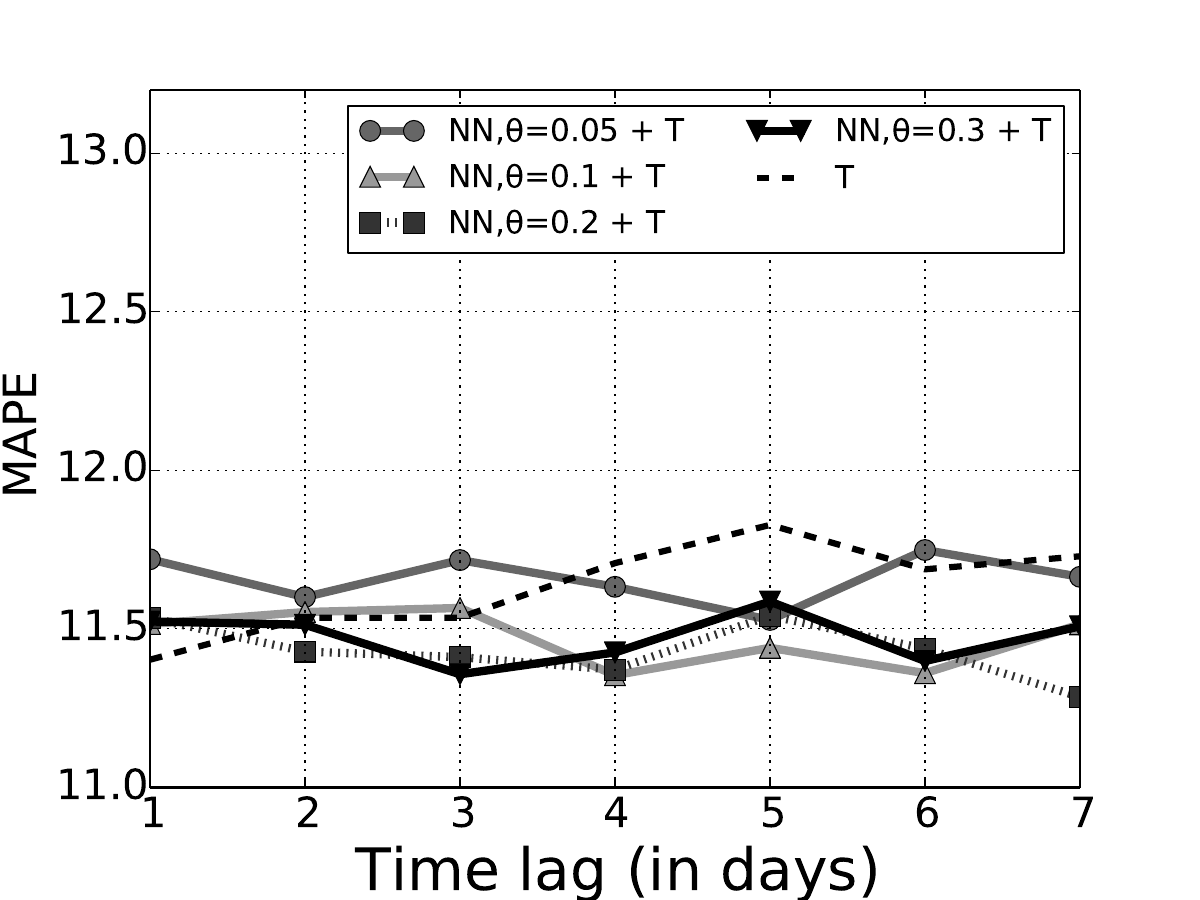}}
~~~~\subfigure[Performance of combined NN+T method (continuous line) vs. topic-based method alone (dashed line)\label{fig:article-pred-NN_plus_T}]{\includegraphics[width=.31\textwidth]{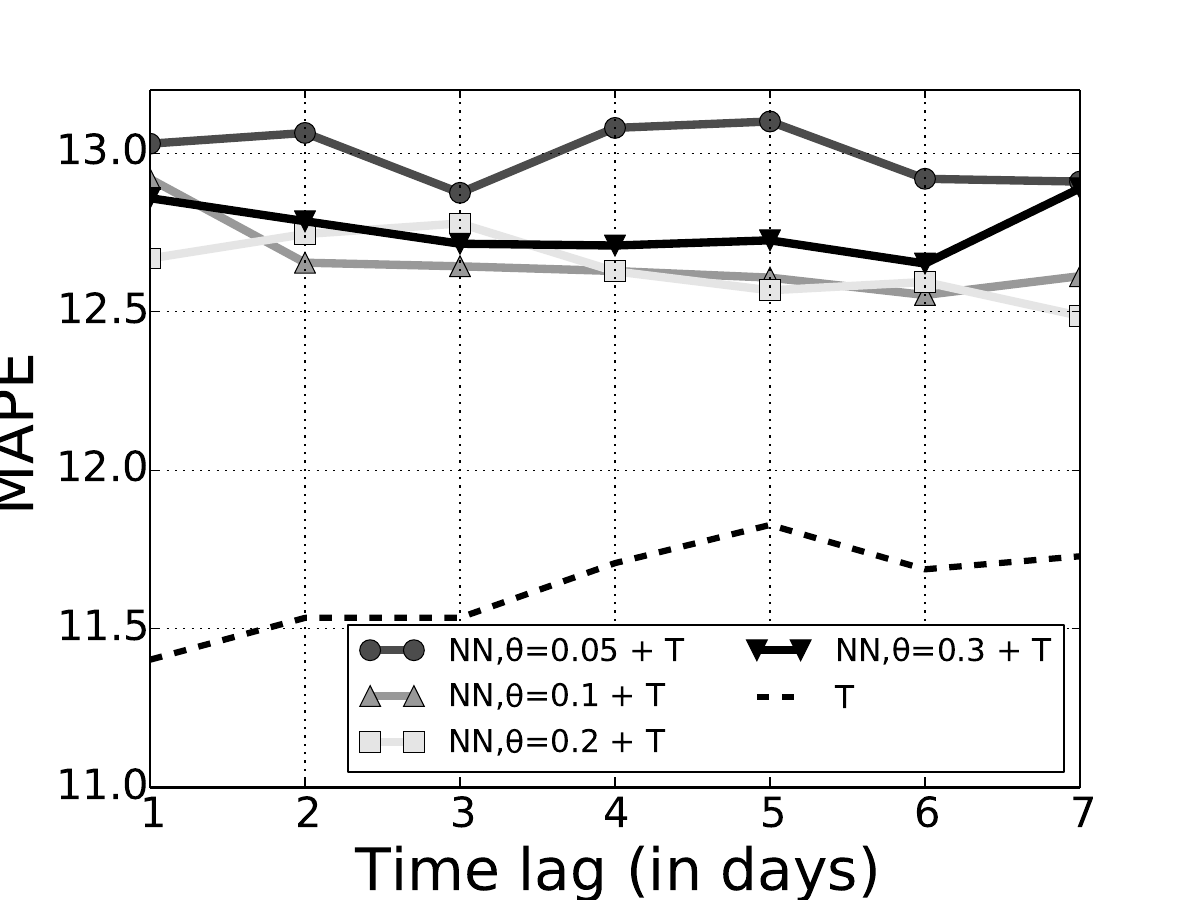}}
~~~~\subfigure[Comparison of NN, T, and NN+T with a fixed $\theta=0.2$\label{fig:article-pred-NN_plus_T-comparison}]{\includegraphics[width=.31\textwidth]{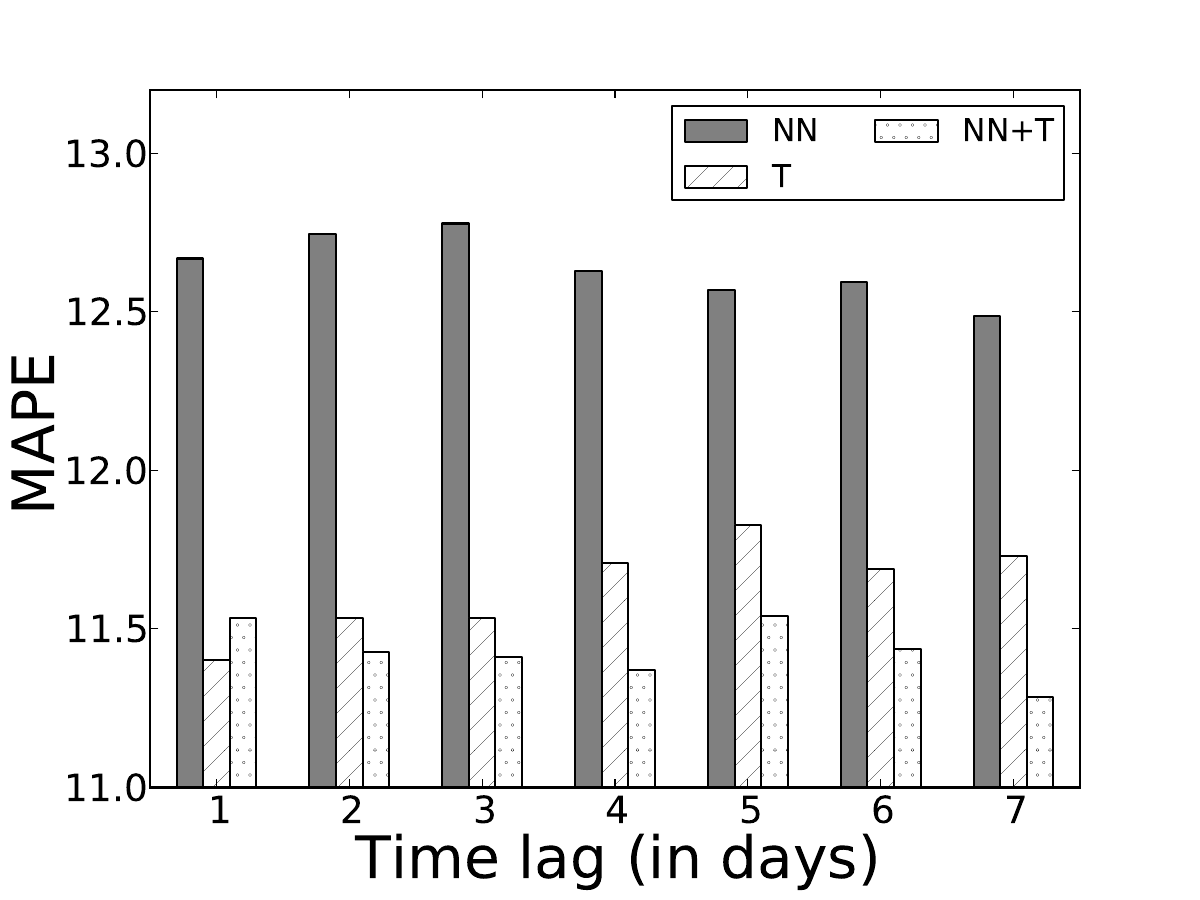}}
\caption{Article popularity prediction using the nearest-neighbors method (NN), the topic-based method (T), and a combined method (NN+T). The first two plots vary $\theta \in \{ 0.05, 0.1, 0.2, 0.3\}$. The last plot fixes $\theta = 0.2$.}
\label{fig:article-pred-results-nn-t}
\end{figure*}

\section{Article predictions}\label{sec:article-predictions}

We now address the problem of predicting the number of visits to an \textit{article}.
We predict the number of cumulative visits to an article $a$ during its first $h$ days after publication, which we denote as $\articlecumvisits{a}{t_a + h}$. As a conservative setting considering the effective half-life measured on Section~\ref{subsec:data-lifespan}, we set $h=3$ days.

Our objective is to assess to which extent topicality and article similarity can help predict the number of visits an article will receive. We start by computing the popularity of a news article as a function of the popularity that similar articles have attained in the last few days (Section~\ref{subsec:art-nn}). Then, we present a method that complements this approach with information about topic popularity in the last few days (Section~\ref{subsec:art-observed-topic}).
Next, we integrate topic popularity predictions into the overall forecasting model to provide (plausible) knowledge about popularity in the future (Section~\ref{subsec:art-predicted-topic}).
Finally, we complement our prediction with early traffic observations to improve over both methods (Section~\ref{subsec:art-early}).

\subsection{Prediction Based on Article Similarity\\Using Nearest Neighbors (NN)}\label{subsec:art-nn}

We hypothesize that similar articles posted within a relatively small time window receive a similar number of visits. The rationale behind this hypothesis is that people who visited an article about a developing story yesterday (or a few days ago), are likely to visit similar articles published today or at a later day. Sets of follow-up articles can be understood as playing the role of ephemeral pseudo-topics.

We measure article similarity by representing articles \articles using $\tfidf$ vectors over the concatenation of their content and title. The similarity between each pair of articles is measured using cosine similarity $\articlesim{\cdot}{\cdot} \in [0,1]$.

To predict article visits, we use these similarities as input to a nearest-neighbors estimation method (NN). This method consists on estimating the value of a function at given point, as an aggregate of the value of that function for a set of points near it~\cite{atkeson_1997_locally,navot_2006_nn}.
We use a variant of the kNN method applied to popularity prediction by~\citet{li2013popularity}, where the number of views of an item is the weighted sum of the number of views of similar items in the past few days.

Given an article $a$ posted on day $t_a$, and a similarity threshold $\theta$, we define $\nnd{\theta}{t}{a}$ as the set of articles published on day $t$ whose similarity with $a$ is greater than or equal to~$\theta$:
\begin{equation}
\nnd{\theta}{t}{a}=\{b \in \articles, \articlesim{a}{b} \geq \theta \wedge t_b = t \}~. 
\end{equation}

We next define a function which gives the weighted average of the number of visits to articles in $\nnd{\theta}{t}{a}$ (for $t < t_a$) up to date $t_a$:
\begin{equation}\label{eq:xa}
X_a(t) = \sum_{b \in \nnd{\theta}{t}{a} } \frac{  \articlesim{a}{b} \cdot \articlecumvisits{b}{t_a} }{ \sum_{b \in \nnd{\theta}{t}{a}  } \articlecumvisits{b}{t_a} }
\end{equation}
\noindent where $\articlecumvisits{b}{t_a}$ is the cumulative number of visits received by article $b$ from  its publication up to and including the publication date of $a$, $t_a$. Finally, our estimator is based on linear regression:
\begin{equation}
\articlecumvisitsest{a}{t_a+h} = \alpha_i + \sum_{i \in \timelags{t_a}} \beta_i X_a(i) + \varepsilon
\end{equation}

\noindent where as before $\timelags{t_a} = \{ t_a - \delta, t_a - \delta + 1, \dots, t_a - 1 \}$ is the set of time lags under consideration, $\alpha$ and $\beta_i$ are the linear regression coefficients, and $\varepsilon$ is the residual term. 

Results are shown on Figure~\ref{fig:article-pred-NN}. The model is trained on 80\% of the data, and tested on the remaining 20\%. We vary $\delta$ from 1 to 7 days and set $\theta$ to values in $\{ 0.05, 0.1, 0.2, 0.3 \}$. We observe that adding more days does not improve significantly the results.
Values of $\theta$ close to 0.1 and 0.2 yield in general better results than 0.05 (which may cover too many articles distantly related to the one for which the prediction is being done) or 0.3 (which may be too strict as a criterion and include too few neighbors).
We experimented with SVR and found the results to be no better than those obtained with linear regression (LR); in the remainder we report only the results with LR which is a simpler model.

\subsection{Prediction Based on Topic Volume (NN+T)}\label{subsec:art-observed-topic}

Let us now consider a predictor of visits to article $a$ based on the topic volume of its main topic $u_a$. This predictor is simply:
\begin{equation}
\articlecumvisitsest{a}{t_a+h} = \alpha_i + \sum_{i \in \timelags{t_a}} \beta_i \topicvisits{u_a}{i} + \varepsilon
\end{equation}
\noindent where $\topicvisits{u_a}{i}$ is the number of visits to topic $u_a$ at time $i$. The result is the dashed line in Figure~\ref{fig:article-pred-NN}. We observe its MAPE value is 1.33 percentage points lower than the one obtained with the method based on NN.

Given that this method is complementary to the one using nearest neighbors, we can combine them using:
\begin{equation}
\articlecumvisitsest{a}{t_a+h} = \alpha_i + \sum_{i \in \timelags{t_a}} \beta_i X_a(i, t_a) + \sum_{i \in \timelags{t_a}} \gamma_i \topicvisits{u_a}{i} + \varepsilon
\end{equation}
\noindent where $X_a(i, t_a)$ is the aggregate of visits to nearest neighbors defined in Equation~\ref{eq:xa}. 

Results are shown on Figure~\ref{fig:article-pred-NN_plus_T}. We observe that the combined method is better than the method based only on topic volume for $\delta > 1$, and that in general the MAPE for $\delta=3$ or $\delta=4$ is lower than for $\delta=1$.

\subsection{Adding Predicted Topic Volume (NN+T+PT)}\label{subsec:art-predicted-topic}

We further improve the results by creating an \textit{ensemble} forecasting that operates in two steps. First, we predict the future popularity of $a$'s topic $u_a$ at time $t_a + h$, $\topicvisitsest{u_a}{t_a + h}$ using the best estimator from Section~\ref{subsec:topic-results}. Next, we incorporate this as an input variable for the regression:
\begin{eqnarray*}
\articlecumvisitsest{a}{t_a+h} & = & \alpha_i + \sum_{i \in \timelags{t_a}} \beta_i X_a(i, t_a) + \sum_{i \in \timelags{t_a}} \gamma_i \topicvisits{u_a}{i} \\
 & & + \eta \topicvisitsest{u_a}{t_a + h} + \varepsilon
\end{eqnarray*}

\begin{figure}[h!]
\centering
\includegraphics[width=.65\columnwidth]{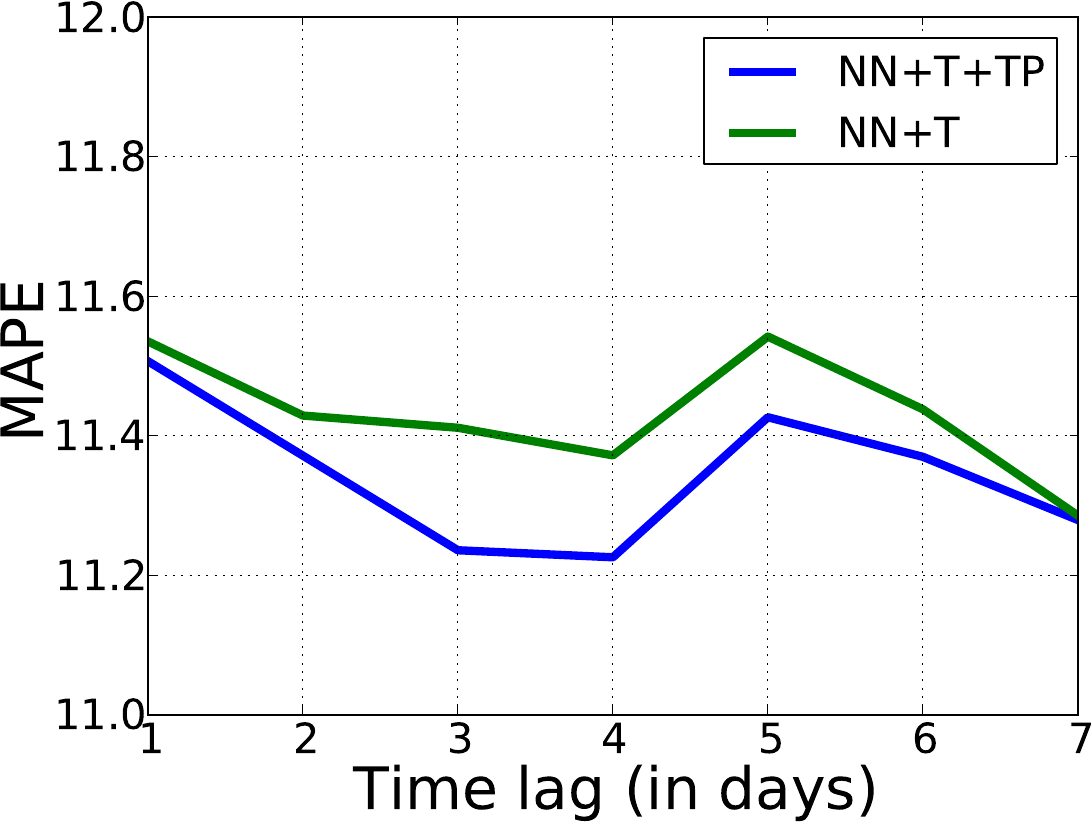}
\caption{Prediction of article visits using nearest neighbors, observed topic volume, and predicted topic volume. The similarity threshold $\theta$ is set to 0.1.}
\label{fig:article-pred-results-tp}
\end{figure}

Results are shown on Figure~\ref{fig:article-pred-results-tp}. We observe a small but consistent improvement when incorporating this variable to our best predictor so far. Again, best results are observed using $\delta=3$ or $\delta=4$.

\subsection{Incorporating Early Observations}\label{subsec:art-early}

\begin{figure}[h!]
\includegraphics[width=0.9\columnwidth]{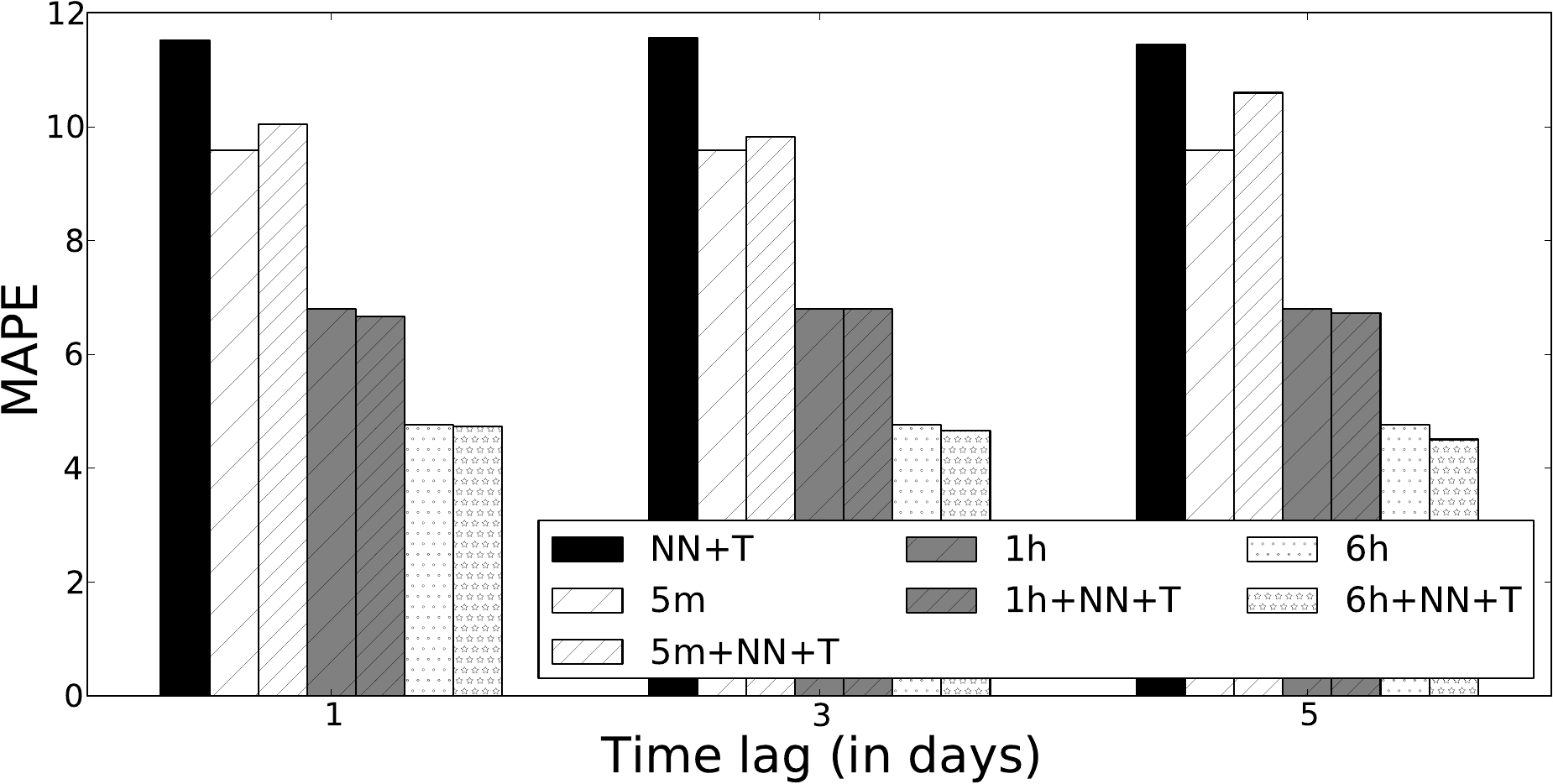}
\caption{Prediction of article visits using our method, compared to methods using early measurements at 5 minutes (5m), 1 hour (1h) and six hours (6h). The similarity threshold $\theta$ is set to 0.1.}
\label{fig:article-pred-results-early1}
\end{figure}

Finally, we compare our method to the standard auto-regressive models based on early measurements (e.g.~\cite{rowe2011forecasting, li2013popularity}).
Results are shown on Figure~\ref{fig:article-pred-results-early1}. We observe that our method yields an error rate on the same scale as methods that use early observations. There is a smooth transition between the error rate resulting from our method (which can be used before publishing the article), and the error rate resulting from methods that use 5 minutes, 1 hour, or 6 hours of early observations. 

On average, our method yields a MAPE of 11.47\%, while early predictions after 5 minutes, 1 hour and six hours obtain error rates of 9.59\%, 6.83\%, and 4.75\% respectively.

In the news domain, it is not realistic that an editor would publish a news article just to verify if it will have a large impact or not. Once a news is published, it can not be withdrawn without a reputational cost. Hence, our method provides a unique competitive advantage over the early-measurements-based methods.

\section{Conclusions and Future Work}

Predicting the popularity of an article before its date of publication requires combining content-based methods, which capture the article's communicative frame, with time series methods, which capture the evolution of people's attention around different issues.
Our approach successfully combines two dimensions in the forecasting of visits for an article: the popularity of similar articles of recent issue, and the popularity of the topics that the article treats.
More specifically, we have shown that an integration of these two dimensions rivals the performance of each dimension on its own. Furthermore, integrating topic predictions---which we can do with as little error as 2.5\%---yield a final mean average error rates of about 11\% when information from the 2 or 3 preceding days is taken into account. 

Next, we plan to use a Content Analysis paradigm to develop a systematic augmentation of the dimensions of article popularity used in this paper. According to~\citet{holsti1969content}, the analysis of a message entails an understanding of who are the source and recipient of the communicative act, what is being said and how, and what are the purpose and potential reach of the message.
In this study we have primarily focused on content and style. In future work, we will integrate information about source, target, purpose (e.g. attitude towards the topic treated) and potential reach (e.g. readability, trust), as well as possible sources of competition for attention (e.g. similar articles on the same day), as a way of increasing the accuracy and robustness of the approach we have presented.

\balance
{
\bibliographystyle{ACM-Reference-Format}
\bibliography{biblio} 


\begin{thebibliography}{00}


\ifx \showCODEN    \undefined \def \showCODEN     #1{\unskip}     \fi
\ifx \showDOI      \undefined \def \showDOI       #1{{\tt DOI:}\penalty0{#1}\ }
  \fi
\ifx \showISBNx    \undefined \def \showISBNx     #1{\unskip}     \fi
\ifx \showISBNxiii \undefined \def \showISBNxiii  #1{\unskip}     \fi
\ifx \showISSN     \undefined \def \showISSN      #1{\unskip}     \fi
\ifx \showLCCN     \undefined \def \showLCCN      #1{\unskip}     \fi
\ifx \shownote     \undefined \def \shownote      #1{#1}          \fi
\ifx \showarticletitle \undefined \def \showarticletitle #1{#1}   \fi
\ifx \showURL      \undefined \def \showURL       #1{#1}          \fi
\providecommand\bibfield[2]{#2}
\providecommand\bibinfo[2]{#2}
\providecommand\natexlab[1]{#1}
\providecommand\showeprint[2][]{arXiv:#2}

\bibitem[\protect\citeauthoryear{Ahmed, Spagna, Huici, and Niccolini}{Ahmed
  et~al\mbox{.}}{2013}]%
        {Ahmed:2013}
\bibfield{author}{\bibinfo{person}{Mohamed Ahmed}, \bibinfo{person}{Stella
  Spagna}, \bibinfo{person}{Felipe Huici}, {and} \bibinfo{person}{Saverio
  Niccolini}.} \bibinfo{year}{2013}\natexlab{}.
\newblock \showarticletitle{A Peek into the Future: Predicting the Evolution of
  Popularity in User Generated Content}. In \bibinfo{booktitle}{{\em Proc. of
  {WSDM}}}. \bibinfo{publisher}{ACM}, \bibinfo{address}{Rome, Italy},
  \bibinfo{pages}{607--616}.
\newblock
\showISBNx{978-1-4503-1869-3}
\showDOI{%
\url{http://dx.doi.org/10.1145/2433396.2433473}}


\bibitem[\protect\citeauthoryear{Armstrong and Collopy}{Armstrong and
  Collopy}{1992}]%
        {armstrong1992error}
\bibfield{author}{\bibinfo{person}{J~Scott Armstrong} {and}
  \bibinfo{person}{Fred Collopy}.} \bibinfo{year}{1992}\natexlab{}.
\newblock \showarticletitle{Error measures for generalizing about forecasting
  methods: Empirical comparisons}.
\newblock \bibinfo{journal}{{\em International Journal of Forecasting\/}}
  \bibinfo{volume}{8}, \bibinfo{number}{1} (\bibinfo{year}{1992}),
  \bibinfo{pages}{69--80}.
\newblock


\bibitem[\protect\citeauthoryear{Arun, Suresh, Veni~Madhavan, and
  Narasimha~Murthy}{Arun et~al\mbox{.}}{2010}]%
        {Arun:2010}
\bibfield{author}{\bibinfo{person}{R. Arun}, \bibinfo{person}{V. Suresh},
  \bibinfo{person}{C.~E. Veni~Madhavan}, {and} \bibinfo{person}{M.~N.
  Narasimha~Murthy}.} \bibinfo{year}{2010}\natexlab{}.
\newblock \showarticletitle{On Finding the Natural Number of Topics with Latent
  Dirichlet Allocation: Some Observations}. In \bibinfo{booktitle}{{\em Proc.
  of PAKDD}}. \bibinfo{publisher}{Springer-Verlag},
  \bibinfo{address}{Hyderabad, India}, \bibinfo{pages}{391--402}.
\newblock
\showISBNx{3-642-13656-7, 978-3-642-13656-6}
\showDOI{%
\url{http://dx.doi.org/10.1007/978-3-642-13657-3_43}}


\bibitem[\protect\citeauthoryear{Atkeson, Moore, and Schaal}{Atkeson
  et~al\mbox{.}}{1997}]%
        {atkeson_1997_locally}
\bibfield{author}{\bibinfo{person}{Christopher~G. Atkeson},
  \bibinfo{person}{Andrew~W. Moore}, {and} \bibinfo{person}{Stefan Schaal}.}
  \bibinfo{year}{1997}\natexlab{}.
\newblock \showarticletitle{Locally Weighted Learning}.
\newblock \bibinfo{journal}{{\em Artificial Intelligence Review\/}}
  \bibinfo{volume}{11}, \bibinfo{number}{1-5} (\bibinfo{year}{1997}),
  \bibinfo{pages}{11--73}.
\newblock
\showURL{%
\url{http://citeseer.ist.psu.edu/atkeson96locally.html}}


\bibitem[\protect\citeauthoryear{Bandari, Asur, and Huberman}{Bandari
  et~al\mbox{.}}{2012}]%
        {Bandari:2012}
\bibfield{author}{\bibinfo{person}{Roja Bandari}, \bibinfo{person}{Sitaram
  Asur}, {and} \bibinfo{person}{Bernardo~A. Huberman}.}
  \bibinfo{year}{2012}\natexlab{}.
\newblock \showarticletitle{The Pulse of News in Social Media: Forecasting
  Popularity}. In \bibinfo{booktitle}{{\em Proc. of {ICWSM}}}.
\newblock


\bibitem[\protect\citeauthoryear{Berger and Milkman}{Berger and
  Milkman}{2012}]%
        {Berger:2012}
\bibfield{author}{\bibinfo{person}{Jonah Berger} {and}
  \bibinfo{person}{Katherine~L. Milkman}.} \bibinfo{year}{2012}\natexlab{}.
\newblock \showarticletitle{What Makes Online Content Viral?}
\newblock \bibinfo{journal}{{\em Journal of Marketing Research\/}}
  \bibinfo{volume}{49}, \bibinfo{number}{2} (\bibinfo{date}{April}
  \bibinfo{year}{2012}), \bibinfo{pages}{192--205}.
\newblock


\bibitem[\protect\citeauthoryear{{Bitly Science Team}}{{Bitly Science
  Team}}{2011}]%
        {bitly_2011_halflife}
\bibfield{author}{\bibinfo{person}{{Bitly Science Team}}.}
  \bibinfo{year}{2011}\natexlab{}.
\newblock \bibinfo{title}{You just shared a link. How long will people pay
  attention?}
\newblock \bibinfo{howpublished}{The {Bit.ly} blog}.   (\bibinfo{date}{6 Sept.}
  \bibinfo{year}{2011}).
\newblock
\showURL{%
\url{http://blog.bitly.com/post/9887686919/you-just-shared-a-link-how-long-will-people-pay}}


\bibitem[\protect\citeauthoryear{{Bitly Science Team}}{{Bitly Science
  Team}}{2012}]%
        {bitly_2012_halflife}
\bibfield{author}{\bibinfo{person}{{Bitly Science Team}}.}
  \bibinfo{year}{2012}\natexlab{}.
\newblock \bibinfo{title}{Halflife by topic}.
\newblock \bibinfo{howpublished}{The {Bit.ly} blog}.   (\bibinfo{date}{9 Nov.}
  \bibinfo{year}{2012}).
\newblock
\showURL{%
\url{http://blog.bitly.com/post/35341087592/halflife-by-topic}}


\bibitem[\protect\citeauthoryear{Blei, Ng, and Jordan}{Blei
  et~al\mbox{.}}{2003}]%
        {Blei:2003}
\bibfield{author}{\bibinfo{person}{David~M. Blei}, \bibinfo{person}{Andrew~Y.
  Ng}, {and} \bibinfo{person}{Michael~I. Jordan}.}
  \bibinfo{year}{2003}\natexlab{}.
\newblock \showarticletitle{Latent Dirichlet Allocation}.
\newblock \bibinfo{journal}{{\em J. Mach. Learn. Res.\/}}  \bibinfo{volume}{3}
  (\bibinfo{date}{March} \bibinfo{year}{2003}), \bibinfo{pages}{993--1022}.
\newblock
\showISSN{1532-4435}
\showURL{%
\url{http://dl.acm.org/citation.cfm?id=944919.944937}}


\bibitem[\protect\citeauthoryear{Castillo, El-Haddad, Pfeffer, and
  Stempeck}{Castillo et~al\mbox{.}}{2014}]%
        {Castillo:2014}
\bibfield{author}{\bibinfo{person}{Carlos Castillo}, \bibinfo{person}{Mohammed
  El-Haddad}, \bibinfo{person}{J\"{u}rgen Pfeffer}, {and} \bibinfo{person}{Matt
  Stempeck}.} \bibinfo{year}{2014}\natexlab{}.
\newblock \showarticletitle{Characterizing the Life Cycle of Online News
  Stories Using Social Media Reactions}. In \bibinfo{booktitle}{{\em Proc. of
  {CSCW}}}. \bibinfo{publisher}{ACM}, \bibinfo{address}{Baltimore, Maryland,
  USA}, \bibinfo{pages}{211--223}.
\newblock
\showISBNx{978-1-4503-2540-0}
\showDOI{%
\url{http://dx.doi.org/10.1145/2531602.2531623}}


\bibitem[\protect\citeauthoryear{Caumont}{Caumont}{2013}]%
        {pew2013trends}
\bibfield{author}{\bibinfo{person}{Andrea Caumont}.}
  \bibinfo{year}{2013}\natexlab{}.
\newblock \bibinfo{title}{12 trends shaping digital news}.
\newblock
  \bibinfo{howpublished}{\url{http://www.pewresearch.org/fact-tank/2013/10/16/12-trends-shaping-digital-news/}}.
    (\bibinfo{date}{16 Oct.} \bibinfo{year}{2013}).
\newblock


\bibitem[\protect\citeauthoryear{Cheng, Adamic, Dow, Kleinberg, and
  Leskovec}{Cheng et~al\mbox{.}}{2014}]%
        {Cheng:2014}
\bibfield{author}{\bibinfo{person}{Justin Cheng}, \bibinfo{person}{Lada~A.
  Adamic}, \bibinfo{person}{P.~Alex Dow}, \bibinfo{person}{Jon~M. Kleinberg},
  {and} \bibinfo{person}{Jure Leskovec}.} \bibinfo{year}{2014}\natexlab{}.
\newblock \showarticletitle{Can cascades be predicted?}. In
  \bibinfo{booktitle}{{\em Proc. of {WWW}}}. \bibinfo{pages}{925--936}.
\newblock


\bibitem[\protect\citeauthoryear{Dezs\"{o}, Almaas, Luk\'{a}cs, R\'{a}cz,
  Szakad\'{a}t, and Barab\'{a}si}{Dezs\"{o} et~al\mbox{.}}{2006}]%
        {dezso_2006_dynamics}
\bibfield{author}{\bibinfo{person}{Z. Dezs\"{o}}, \bibinfo{person}{E. Almaas},
  \bibinfo{person}{A. Luk\'{a}cs}, \bibinfo{person}{B. R\'{a}cz},
  \bibinfo{person}{I. Szakad\'{a}t}, {and} \bibinfo{person}{A.~L.
  Barab\'{a}si}.} \bibinfo{year}{2006}\natexlab{}.
\newblock \showarticletitle{Dynamics of information access on the web}.
\newblock \bibinfo{journal}{{\em Physical Review E (Statistical, Nonlinear, and
  Soft Matter Physics)\/}} \bibinfo{volume}{73}, \bibinfo{number}{6}
  (\bibinfo{year}{2006}), \bibinfo{pages}{066132+}.
\newblock
\showDOI{%
\url{http://dx.doi.org/10.1103/PhysRevE.73.066132}}
\showeprint[arxiv]{physics/0505087}


\bibitem[\protect\citeauthoryear{Galtung and Ruge}{Galtung and Ruge}{1965}]%
        {galtung_1965_foreign}
\bibfield{author}{\bibinfo{person}{Johan Galtung} {and}
  \bibinfo{person}{Mari~H. Ruge}.} \bibinfo{year}{1965}\natexlab{}.
\newblock \showarticletitle{The Structure of Foreign News}.
\newblock \bibinfo{journal}{{\em Journal of Peace Research\/}}
  \bibinfo{volume}{2}, \bibinfo{number}{1} (\bibinfo{year}{1965}),
  \bibinfo{pages}{64--91}.
\newblock
\showURL{%
\url{http://www.jstor.org/stable/423011}}


\bibitem[\protect\citeauthoryear{Hofmann}{Hofmann}{1999}]%
        {Hofmann:1999}
\bibfield{author}{\bibinfo{person}{Thomas Hofmann}.}
  \bibinfo{year}{1999}\natexlab{}.
\newblock \showarticletitle{Probabilistic Latent Semantic Indexing}. In
  \bibinfo{booktitle}{{\em Proc. of {SIGIR}}}. \bibinfo{pages}{50--57}.
\newblock


\bibitem[\protect\citeauthoryear{Holsti}{Holsti}{1969}]%
        {holsti1969content}
\bibfield{author}{\bibinfo{person}{Ole~R Holsti}.}
  \bibinfo{year}{1969}\natexlab{}.
\newblock \bibinfo{booktitle}{{\em Content analysis for the social sciences and
  humanities}}.
\newblock \bibinfo{publisher}{Addison-Wesley Reading, MA}.
\newblock


\bibitem[\protect\citeauthoryear{Hsieh, Moghbel, Fang, and Cho}{Hsieh
  et~al\mbox{.}}{2013}]%
        {Hsieh:2013}
\bibfield{author}{\bibinfo{person}{C Hsieh}, \bibinfo{person}{Christopher
  Moghbel}, \bibinfo{person}{Jianhong Fang}, {and} \bibinfo{person}{Junghoo
  Cho}.} \bibinfo{year}{2013}\natexlab{}.
\newblock \bibinfo{booktitle}{{\em Experts vs the crowd: Examining popular news
  prediction perfomance on {Twitter}}}.
\newblock \bibinfo{type}{{T}echnical {R}eport}. \bibinfo{institution}{UCLA}.
\newblock


\bibitem[\protect\citeauthoryear{Huang, Chen, Luo, and Lee}{Huang
  et~al\mbox{.}}{2012}]%
        {huang_2012_predicting}
\bibfield{author}{\bibinfo{person}{Shu Huang}, \bibinfo{person}{Min Chen},
  \bibinfo{person}{Bo Luo}, {and} \bibinfo{person}{Dongwon Lee}.}
  \bibinfo{year}{2012}\natexlab{}.
\newblock \showarticletitle{Predicting Aggregate Social Activities Using
  {Continuous-Time} Stochastic Process}. In \bibinfo{booktitle}{{\em Proc. of
  CIKM 2012}}. \bibinfo{address}{Maui, Hawaii, USA}.
\newblock


\bibitem[\protect\citeauthoryear{Jamali and Rangwala}{Jamali and
  Rangwala}{2009}]%
        {Jamali:2009}
\bibfield{author}{\bibinfo{person}{Salman Jamali} {and} \bibinfo{person}{Huzefa
  Rangwala}.} \bibinfo{year}{2009}\natexlab{}.
\newblock \showarticletitle{Digging Digg: Comment Mining, Popularity
  Prediction, and Social Network Analysis}. In \bibinfo{booktitle}{{\em Proc.
  of {WISM}}}. \bibinfo{publisher}{IEEE Computer Society},
  \bibinfo{address}{Washington, DC, USA}, \bibinfo{pages}{32--38}.
\newblock
\showISBNx{978-0-7695-3817-4}
\showDOI{%
\url{http://dx.doi.org/10.1109/WISM.2009.15}}


\bibitem[\protect\citeauthoryear{Kim, Kim, and Cho}{Kim et~al\mbox{.}}{2011}]%
        {kim_2011_temperature}
\bibfield{author}{\bibinfo{person}{Su-Do Kim}, \bibinfo{person}{Sung-Hwan Kim},
  {and} \bibinfo{person}{Hwan-Gue Cho}.} \bibinfo{year}{2011}\natexlab{}.
\newblock \showarticletitle{Predicting the Virtual Temperature of {Web-Blog}
  Articles as a Measurement Tool for Online Popularity}. In
  \bibinfo{booktitle}{{\em Proc. of {CIT}}}. \bibinfo{publisher}{IEEE},
  \bibinfo{pages}{449--454}.
\newblock
\showISBNx{978-1-4577-0383-6}
\showDOI{%
\url{http://dx.doi.org/10.1109/CIT.2011.104}}


\bibitem[\protect\citeauthoryear{Kwak, Lee, Park, and Moon}{Kwak
  et~al\mbox{.}}{2010}]%
        {kwak:2010}
\bibfield{author}{\bibinfo{person}{Haewoon Kwak}, \bibinfo{person}{Changhyun
  Lee}, \bibinfo{person}{Hosung Park}, {and} \bibinfo{person}{Sue Moon}.}
  \bibinfo{year}{2010}\natexlab{}.
\newblock \showarticletitle{What is {Twitter}, a Social Network or a News
  Media?}. In \bibinfo{booktitle}{{\em Proc. of {WWW}}}.
  \bibinfo{publisher}{ACM}, \bibinfo{address}{Raleigh, North Carolina, USA},
  \bibinfo{pages}{591--600}.
\newblock
\showISBNx{978-1-60558-799-8}
\showDOI{%
\url{http://dx.doi.org/10.1145/1772690.1772751}}


\bibitem[\protect\citeauthoryear{Lakkaraju and Ajmera}{Lakkaraju and
  Ajmera}{2011}]%
        {lakkaraju_2011_attention}
\bibfield{author}{\bibinfo{person}{Himabindu Lakkaraju} {and}
  \bibinfo{person}{Jitendra Ajmera}.} \bibinfo{year}{2011}\natexlab{}.
\newblock \showarticletitle{Attention prediction on social media brand pages}.
  In \bibinfo{booktitle}{{\em Proc. of CIKM}}. \bibinfo{publisher}{ACM},
  \bibinfo{address}{Glasgow, Scotland, UK}, \bibinfo{pages}{2157--2160}.
\newblock
\showISBNx{978-1-4503-0717-8}
\showDOI{%
\url{http://dx.doi.org/10.1145/2063576.2063915}}


\bibitem[\protect\citeauthoryear{Lakkaraju, McAuley, and Leskovec}{Lakkaraju
  et~al\mbox{.}}{2013}]%
        {lakkaraju_2013_reddit}
\bibfield{author}{\bibinfo{person}{Himabindu Lakkaraju},
  \bibinfo{person}{Julian~J. McAuley}, {and} \bibinfo{person}{Jure Leskovec}.}
  \bibinfo{year}{2013}\natexlab{}.
\newblock \showarticletitle{What's in a Name? Understanding the Interplay
  between Titles, Content, and Communities in Social Media}. In
  \bibinfo{booktitle}{{\em Proc. of ICWSM}}. \bibinfo{publisher}{AAAI Press}.
\newblock


\bibitem[\protect\citeauthoryear{Lee and Seung}{Lee and Seung}{2000}]%
        {Lee:2000}
\bibfield{author}{\bibinfo{person}{Daniel~D. Lee} {and}
  \bibinfo{person}{H.~Sebastian Seung}.} \bibinfo{year}{2000}\natexlab{}.
\newblock \showarticletitle{Algorithms for Non-negative Matrix Factorization}.
  In \bibinfo{booktitle}{{\em In NIPS}}. \bibinfo{publisher}{MIT Press},
  \bibinfo{pages}{556--562}.
\newblock


\bibitem[\protect\citeauthoryear{Lee, Moon, and Salamatian}{Lee
  et~al\mbox{.}}{2010}]%
        {lee_2010_popularity}
\bibfield{author}{\bibinfo{person}{Jong~G. Lee}, \bibinfo{person}{Sue Moon},
  {and} \bibinfo{person}{Kave Salamatian}.} \bibinfo{year}{2010}\natexlab{}.
\newblock \showarticletitle{An Approach to Model and Predict the Popularity of
  Online Contents with Explanatory Factors}. In \bibinfo{booktitle}{{\em IEEE
  Conference on Web Intelligence}}. \bibinfo{address}{Toronto, Canada}.
\newblock


\bibitem[\protect\citeauthoryear{Lerman and Hogg}{Lerman and Hogg}{2010a}]%
        {Lerman:2010}
\bibfield{author}{\bibinfo{person}{Kristina Lerman} {and} \bibinfo{person}{Tad
  Hogg}.} \bibinfo{year}{2010}\natexlab{a}.
\newblock \showarticletitle{Using a Model of Social Dynamics to Predict
  Popularity of News}. In \bibinfo{booktitle}{{\em Proc. of {WWW}}}.
  \bibinfo{publisher}{ACM}, \bibinfo{address}{Raleigh, North Carolina, USA},
  \bibinfo{pages}{621--630}.
\newblock
\showISBNx{978-1-60558-799-8}
\showDOI{%
\url{http://dx.doi.org/10.1145/1772690.1772754}}


\bibitem[\protect\citeauthoryear{Lerman and Hogg}{Lerman and Hogg}{2010b}]%
        {lerman_2010_news}
\bibfield{author}{\bibinfo{person}{Kristina Lerman} {and} \bibinfo{person}{Tad
  Hogg}.} \bibinfo{year}{2010}\natexlab{b}.
\newblock \showarticletitle{Using a model of social dynamics to predict
  popularity of news}. In \bibinfo{booktitle}{{\em Proc. of {WWW}}}.
  \bibinfo{publisher}{ACM}, \bibinfo{address}{Raleigh, North Carolina, USA},
  \bibinfo{pages}{621--630}.
\newblock
\showISBNx{978-1-60558-799-8}
\showDOI{%
\url{http://dx.doi.org/10.1145/1772690.1772754}}


\bibitem[\protect\citeauthoryear{Li, Ma, Wang, Liu, and Xu}{Li
  et~al\mbox{.}}{2013}]%
        {li2013popularity}
\bibfield{author}{\bibinfo{person}{Haitao Li}, \bibinfo{person}{Xiaoqiang Ma},
  \bibinfo{person}{Feng Wang}, \bibinfo{person}{Jiangchuan Liu}, {and}
  \bibinfo{person}{Ke Xu}.} \bibinfo{year}{2013}\natexlab{}.
\newblock \showarticletitle{On popularity prediction of videos shared in online
  social networks}. In \bibinfo{booktitle}{{\em Proc. of {CIKM}}}. ACM,
  \bibinfo{pages}{169--178}.
\newblock


\bibitem[\protect\citeauthoryear{McCallum and Nigam}{McCallum and
  Nigam}{1998}]%
        {McCallum:1998}
\bibfield{author}{\bibinfo{person}{Andrew McCallum} {and}
  \bibinfo{person}{Kamal Nigam}.} \bibinfo{year}{1998}\natexlab{}.
\newblock \showarticletitle{A comparison of event models for Naive Bayes text
  classification}. In \bibinfo{booktitle}{{\em IN AAAI-98 WORKSHOP ON LEARNING
  FOR TEXT CATEGORIZATION}}. \bibinfo{publisher}{AAAI Press},
  \bibinfo{pages}{41--48}.
\newblock


\bibitem[\protect\citeauthoryear{M{\"u}ller, Smola, R{\"a}tsch, Sch{\"o}lkopf,
  Kohlmorgen, and Vapnik}{M{\"u}ller et~al\mbox{.}}{1997}]%
        {muller1997predicting}
\bibfield{author}{\bibinfo{person}{K-R M{\"u}ller}, \bibinfo{person}{Alex~J
  Smola}, \bibinfo{person}{Gunnar R{\"a}tsch}, \bibinfo{person}{Bernhard
  Sch{\"o}lkopf}, \bibinfo{person}{Jens Kohlmorgen}, {and}
  \bibinfo{person}{Vladimir Vapnik}.} \bibinfo{year}{1997}\natexlab{}.
\newblock \showarticletitle{Predicting time series with support vector
  machines}.
\newblock In \bibinfo{booktitle}{{\em Artificial Neural Networks ({ICANN})}}.
  \bibinfo{publisher}{Springer}, \bibinfo{pages}{999--1004}.
\newblock


\bibitem[\protect\citeauthoryear{Myers, Zhu, and Leskovec}{Myers
  et~al\mbox{.}}{2012}]%
        {myers_2012_external}
\bibfield{author}{\bibinfo{person}{Seth~A. Myers}, \bibinfo{person}{Chenguang
  Zhu}, {and} \bibinfo{person}{Jure Leskovec}.}
  \bibinfo{year}{2012}\natexlab{}.
\newblock \showarticletitle{Information diffusion and external influence in
  networks}. In \bibinfo{booktitle}{{\em Proc. of {KDD}}}.
  \bibinfo{publisher}{ACM}, \bibinfo{address}{Beijing, China},
  \bibinfo{pages}{33--41}.
\newblock
\showISBNx{978-1-4503-1462-6}
\showDOI{%
\url{http://dx.doi.org/10.1145/2339530.2339540}}


\bibitem[\protect\citeauthoryear{Navot, Shpigelman, Tishby, and Vaadia}{Navot
  et~al\mbox{.}}{2006}]%
        {navot_2006_nn}
\bibfield{author}{\bibinfo{person}{A. Navot}, \bibinfo{person}{L. Shpigelman},
  \bibinfo{person}{N. Tishby}, {and} \bibinfo{person}{Vaadia}.}
  \bibinfo{year}{2006}\natexlab{}.
\newblock \showarticletitle{Nearest neighbor based feature selection for
  regression and its application to neural activity.}. In
  \bibinfo{booktitle}{{\em Proc. of {NIPS}}}.
\newblock


\bibitem[\protect\citeauthoryear{Paice}{Paice}{1990}]%
        {Paice:1990}
\bibfield{author}{\bibinfo{person}{Chris~D. Paice}.}
  \bibinfo{year}{1990}\natexlab{}.
\newblock \showarticletitle{Another Stemmer}.
\newblock \bibinfo{journal}{{\em SIGIR Forum\/}} \bibinfo{volume}{24},
  \bibinfo{number}{3} (\bibinfo{date}{Nov.} \bibinfo{year}{1990}),
  \bibinfo{pages}{56--61}.
\newblock
\showISSN{0163-5840}
\showDOI{%
\url{http://dx.doi.org/10.1145/101306.101310}}


\bibitem[\protect\citeauthoryear{Pinto, Almeida, and Gon\c{c}alves}{Pinto
  et~al\mbox{.}}{2013}]%
        {Pinto:2013}
\bibfield{author}{\bibinfo{person}{Henrique Pinto}, \bibinfo{person}{Jussara~M.
  Almeida}, {and} \bibinfo{person}{Marcos~A. Gon\c{c}alves}.}
  \bibinfo{year}{2013}\natexlab{}.
\newblock \showarticletitle{Using Early View Patterns to Predict the Popularity
  of {YouTube} Videos}. In \bibinfo{booktitle}{{\em Proc. of {WSDM}}}.
  \bibinfo{publisher}{ACM}, \bibinfo{address}{Rome, Italy},
  \bibinfo{pages}{365--374}.
\newblock
\showISBNx{978-1-4503-1869-3}
\showDOI{%
\url{http://dx.doi.org/10.1145/2433396.2433443}}


\bibitem[\protect\citeauthoryear{Rowe}{Rowe}{2011}]%
        {rowe2011forecasting}
\bibfield{author}{\bibinfo{person}{Matthew Rowe}.}
  \bibinfo{year}{2011}\natexlab{}.
\newblock \showarticletitle{Forecasting audience increase on {YouTube}}. In
  \bibinfo{booktitle}{{\em Workshop on User Profile Data on the Social Semantic
  Web}}. \bibinfo{address}{Heraklion, Greece}.
\newblock


\bibitem[\protect\citeauthoryear{Ruan, Purohit, Fuhry, Parthasarathy, and
  Sheth}{Ruan et~al\mbox{.}}{2012}]%
        {ruan_2012_prediction}
\bibfield{author}{\bibinfo{person}{Yiye Ruan}, \bibinfo{person}{Hemant
  Purohit}, \bibinfo{person}{David Fuhry}, \bibinfo{person}{Srinivasan
  Parthasarathy}, {and} \bibinfo{person}{Amit Sheth}.}
  \bibinfo{year}{2012}\natexlab{}.
\newblock \showarticletitle{Prediction of Topic Volume on {Twitter}}. In
  \bibinfo{booktitle}{{\em WebSci (short papers)}}. \bibinfo{address}{Evanston,
  Illinois, USA}.
\newblock
\showURL{%
\url{http://knoesis.org/library/resource.php?id=1698}}


\bibitem[\protect\citeauthoryear{Salganik, Dodds, and Watts}{Salganik
  et~al\mbox{.}}{2006}]%
        {Salganik:2006}
\bibfield{author}{\bibinfo{person}{Matthew~J. Salganik},
  \bibinfo{person}{Peter~Sheridan Dodds}, {and} \bibinfo{person}{Duncan~J.
  Watts}.} \bibinfo{year}{2006}\natexlab{}.
\newblock \showarticletitle{Experimental Study of Inequality and
  Unpredictability in an Artificial Cultural Market}.
\newblock \bibinfo{journal}{{\em Science\/}} \bibinfo{volume}{311},
  \bibinfo{number}{5762} (\bibinfo{year}{2006}), \bibinfo{pages}{854--856}.
\newblock


\bibitem[\protect\citeauthoryear{Smola and Sch{\"o}lkopf}{Smola and
  Sch{\"o}lkopf}{2004}]%
        {smola2004tutorial}
\bibfield{author}{\bibinfo{person}{Alex~J Smola} {and}
  \bibinfo{person}{Bernhard Sch{\"o}lkopf}.} \bibinfo{year}{2004}\natexlab{}.
\newblock \showarticletitle{A tutorial on support vector regression}.
\newblock \bibinfo{journal}{{\em Statistics and computing\/}}
  \bibinfo{volume}{14}, \bibinfo{number}{3} (\bibinfo{year}{2004}),
  \bibinfo{pages}{199--222}.
\newblock


\bibitem[\protect\citeauthoryear{Szabo and Huberman}{Szabo and
  Huberman}{2010a}]%
        {Szabo:2010}
\bibfield{author}{\bibinfo{person}{Gabor Szabo} {and}
  \bibinfo{person}{Bernardo~A. Huberman}.} \bibinfo{year}{2010}\natexlab{a}.
\newblock \showarticletitle{Predicting the Popularity of Online Content}.
\newblock \bibinfo{journal}{{\em Commun. ACM\/}} \bibinfo{volume}{53},
  \bibinfo{number}{8} (\bibinfo{date}{Aug.} \bibinfo{year}{2010}),
  \bibinfo{pages}{80--88}.
\newblock
\showISSN{0001-0782}
\showDOI{%
\url{http://dx.doi.org/10.1145/1787234.1787254}}


\bibitem[\protect\citeauthoryear{Szabo and Huberman}{Szabo and
  Huberman}{2010b}]%
        {szabo2012predicting}
\bibfield{author}{\bibinfo{person}{Gabor Szabo} {and}
  \bibinfo{person}{Bernardo~A. Huberman}.} \bibinfo{year}{2010}\natexlab{b}.
\newblock \showarticletitle{Predicting the popularity of online content}.
\newblock \bibinfo{journal}{{\it Commun. ACM}} \bibinfo{volume}{53},
  \bibinfo{number}{8} (\bibinfo{date}{Aug.} \bibinfo{year}{2010}),
  \bibinfo{pages}{80--88}.
\newblock
\showISSN{0001-0782}
\showDOI{%
\url{http://dx.doi.org/10.1145/1787234.1787254}}


\bibitem[\protect\citeauthoryear{Tatar, Antoniadis, de~Amorim, and Fdida}{Tatar
  et~al\mbox{.}}{2012}]%
        {Tatar:2012}
\bibfield{author}{\bibinfo{person}{Alexandru Tatar}, \bibinfo{person}{Panayotis
  Antoniadis}, \bibinfo{person}{Marcelo~Dias de Amorim}, {and}
  \bibinfo{person}{Serge Fdida}.} \bibinfo{year}{2012}\natexlab{}.
\newblock \showarticletitle{Ranking News Articles Based on Popularity
  Prediction}. In \bibinfo{booktitle}{{\em ASONAM}}. \bibinfo{pages}{106--110}.
\newblock


\bibitem[\protect\citeauthoryear{Tatar, Leguay, Antoniadis, Limbourg,
  de~Amorim, and Fdida}{Tatar et~al\mbox{.}}{2011}]%
        {tatar_2011_predicting}
\bibfield{author}{\bibinfo{person}{Alexandru Tatar},
  \bibinfo{person}{J{\'{e}}r{\'{e}}mie Leguay}, \bibinfo{person}{Panayotis
  Antoniadis}, \bibinfo{person}{Arnaud Limbourg}, \bibinfo{person}{Marcelo~D.
  de Amorim}, {and} \bibinfo{person}{Serge Fdida}.}
  \bibinfo{year}{2011}\natexlab{}.
\newblock \showarticletitle{Predicting the popularity of online articles based
  on user comments}. In \bibinfo{booktitle}{{\em Proc. of {WIMS}}}.
  \bibinfo{publisher}{ACM}, \bibinfo{address}{Sogndal, Norway}.
\newblock
\showISBNx{978-1-4503-0148-0}
\showDOI{%
\url{http://dx.doi.org/10.1145/1988688.1988766}}


\bibitem[\protect\citeauthoryear{Tsagkias, Weerkamp, and De~Rijke}{Tsagkias
  et~al\mbox{.}}{2009}]%
        {tsagkias2009predicting}
\bibfield{author}{\bibinfo{person}{Manos Tsagkias}, \bibinfo{person}{Wouter
  Weerkamp}, {and} \bibinfo{person}{Maarten De~Rijke}.}
  \bibinfo{year}{2009}\natexlab{}.
\newblock \showarticletitle{Predicting the volume of comments on online news
  stories}. In \bibinfo{booktitle}{{\em Proc. of CIKM}}. ACM,
  \bibinfo{pages}{1765--1768}.
\newblock


\bibitem[\protect\citeauthoryear{Tsagkias, Weerkamp, and de~Rijke}{Tsagkias
  et~al\mbox{.}}{2010}]%
        {Tsagkias:2010}
\bibfield{author}{\bibinfo{person}{Manos Tsagkias}, \bibinfo{person}{Wouter
  Weerkamp}, {and} \bibinfo{person}{Maarten de Rijke}.}
  \bibinfo{year}{2010}\natexlab{}.
\newblock \showarticletitle{News Comments: Exploring, Modeling, and Online
  Prediction}. In \bibinfo{booktitle}{{\em Proc. of {ECIR}}}.
  \bibinfo{publisher}{Springer-Verlag}, \bibinfo{address}{Milton Keynes, UK},
  \bibinfo{pages}{191--203}.
\newblock
\showISBNx{3-642-12274-4, 978-3-642-12274-3}
\showDOI{%
\url{http://dx.doi.org/10.1007/978-3-642-12275-0_19}}


\bibitem[\protect\citeauthoryear{Yu, Chen, and Kwok}{Yu et~al\mbox{.}}{2011}]%
        {yu_2011_predicting}
\bibfield{author}{\bibinfo{person}{Bei Yu}, \bibinfo{person}{Miao Chen}, {and}
  \bibinfo{person}{Linchi Kwok}.} \bibinfo{year}{2011}\natexlab{}.
\newblock \showarticletitle{Toward Predicting Popularity of Social Marketing
  Messages}. In \bibinfo{booktitle}{{\em Social Computing, Behavioral-Cultural
  Modeling and Prediction}} {\em (\bibinfo{series}{LNCS})},
  \bibfield{editor}{\bibinfo{person}{J.~Salerno}, \bibinfo{person}{S.~J. Yang},
  \bibinfo{person}{D.~Nau}, {and} \bibinfo{person}{S.~K. Chai}} (Eds.),
  Vol.~\bibinfo{volume}{6589}. \bibinfo{publisher}{Springer},
  \bibinfo{address}{College Park, MD, USA}, \bibinfo{pages}{317--324}.
\newblock


\end{thebibliography}
}

\end{document}